\documentclass[aps,pra,showpacs,superscriptaddress,amssymb,twocolumn]{revtex4-1}
\usepackage{eqnarray,amsmath,braket,physics}
\usepackage{natbib}
\usepackage{graphicx}
\usepackage{enumitem}
\usepackage{appendix}
\usepackage{color}
\usepackage[export]{adjustbox}
\usepackage{epstopdf} 
\DeclareGraphicsExtensions{.pdf,.eps,.png,.jpg,.mps}
\usepackage[colorlinks=true,citecolor=blue,urlcolor=blue,linkcolor=blue]{hyperref}
\usepackage[position=top,singlelinecheck=off,justification=raggedright,caption=false]{subfig}
\usepackage{bm,amsmath}
\usepackage{dcolumn}
\usepackage{multirow}
\usepackage{hyperref}
\usepackage{float}
\newcommand*\diff{\mathop{}\!\mathrm{d}}

\def\BEq{\begin{equation}}
\def\EEq{\end{equation}}
\def\BEqA{\begin{eqnarray}}
\def\EEqA{\end{eqnarray}}
\def\BW{\begin{widetext}}
\def\EW{\end{widetext}}
\usepackage{array}
\usepackage{float}
\usepackage{xfrac}
\begin{document}
\title{Multielectron dots provide faster Rabi oscillations when the core electrons are strongly confined}

\author{H.\ Ekmel Ercan}
%\email{hercan@wisc.edu}
\affiliation{Department of Electrical and Computer Engineering, University of California, Los Angeles, Los Angeles, CA 90095, USA}
\author{Christopher R.\ Anderson}
\affiliation{Department of Mathematics, University of California, Los Angeles, Los Angeles, CA 90095 USA}
\affiliation{Center for Quantum Science and Engineering, University of California, Los Angeles, Los Angeles, CA 90095, USA}
\author{S.\ N.\ Coppersmith}
%\email{snc@physics.wisc.edu}
\affiliation{School of Physics, The University of New South Wales, Sydney, NSW 2052, Australia}
\author{Mark Friesen}
%\email{friesen@physics.wisc.edu}
\affiliation{Department of Physics, University of Wisconsin-Madison, Madison, WI 53706, USA}
\author{Mark F.\ Gyure}
\affiliation{Department of Electrical and Computer Engineering, University of California, Los Angeles, Los Angeles, CA 90095, USA}
\affiliation{Center for Quantum Science and Engineering, University of California, Los Angeles, Los Angeles, CA 90095, USA}
\date{\today}

\begin{abstract}
Increasing the number of electrons in electrostatically confined quantum dots can enable faster qubit gates. Although this has been experimentally demonstrated, a detailed quantitative understanding has been missing. Here we study one- and three-electron quantum dots in silicon/silicon-germanium heterostructures within the context of electrically-driven spin resonance (EDSR) using full configuration interaction and tight binding approaches. Our calculations show that anharmonicity of the confinement potential plays an important role: while the EDSR Rabi frequency of electrons in a harmonic potential is indifferent to the electron number, soft anharmonic confinements lead to larger and hard anharmonic confinements lead to smaller Rabi frequencies. We also confirm that double dots allow fast Rabi oscillations, and further suggest that purposefully engineered confinements can also yield similarly fast Rabi oscillations in a single dot. Finally, we discuss the role of interface steps. These findings have important implications for the design of multielectron Si/SiGe quantum dot qubits.
\end{abstract}    

\maketitle

\section{Introduction}
Electron spin qubits in quantum dots are promising candidates for achieving scalable quantum computing, with recent experiments achieving high-fidelity single and two-qubit gates. While single electron spin control in quantum dot qubits has been achieved via electron spin resonance (ESR)~\cite{Koppens:2006p766,Veldhorst:2014p981}, as the strong oscillating magnetic field requirement of the ESR technique proves to be difficult to achieve on a chip, electrically-driven spin resonance (EDSR) approach has also become a viable alternative~\cite{PioroLadriere:2008p776,Obata:2010p2612,Kawakami:2014p666,Takeda:2016p1600694,Watson:2018p633,Yang:2020p350,Noiri:2022p338}. In EDSR, electrons are spatially oscillated via electric fields in the presence of intrinsic or synthetic spin-orbit coupling (SOC), creating an effective oscillating magnetic field. 
While it is natural to create each qubit using a single electron, any odd number of electrons can potentially be used to encode a two-level low-energy qubit space defined by up and down spins.  In fact, it has been experimentally demonstrated that increasing the number of electrons in a quantum dot can increase the EDSR operating speed and the $Q$-factor of a quantum dot spin qubit in Si~\cite{Leon:2020p797,Leon:2021p3228}. Although it is instructive to draw an analogy with the concept of a single ``valence" electron sitting on an even numbered ``closed-shell" of electrons from chemistry (which we also use in this paper) to understand multielectron single-spin physics, a more precise condition for qubit implementation can simply be described as having a ground state with total spin $S=1/2$ that is Zeeman split by a magnetic field. The  total spin of the ground state of a multielectron system depends sensitively on how the electrons are confined. The shape of the confinement also determines the collective vs.\ relative aspect of the electron motion, which in turn dictates the net spin motion. Therefore, it is clear that understanding the confinement effects quantitatively is highly desirable for multielectron EDSR.

In this paper we focus on EDSR control of three electrons in a quantum dot in a Si/SiGe heterostructure and investigate how the confinement and interface disorder affect its performance in the presence of synthetic SOC. We use the resonant Rabi frequency $f_{Rabi}$ as our performance metric and make comparisons with the single-electron case to find out under which conditions increasing electron number can be desirable.

In recent years, it has become clear that the electron-electron (e-e) interactions in silicon dots can be strong enough to lead to Wigner molecularization, and therefore cannot be described perturbatively in general. Here, we use the full configuration interaction (FCI) method that has been widely used in quantum chemistry and become a prominent approach in the semiconductor quantum dot community to accurately account for strong e-e interactions~\cite{Corrigan:2021p127701,Ercan:2021p235302,Abadillo-Uriel:2021p195305,Yannouleas:2022p21LT01,Yannouleas:2022p205302,Yannouleas:2022preprint}. Another important aspect of quantum dots in Si/SiGe heterostructures is the presence of nearly degenerate conduction band valley states that are highly sensitive to atomistic details of the quantum well interface~\cite{Zwanenburg:2013p961}. These details are accounted for in our computational procedure by using a minimal tight-binding (TB) model to generate the underlying single-electron basis in the FCI calculation.  Finally, we use perturbation theory to include the effects of magnetic fields.

\begin{figure}
\includegraphics[width=3 in]{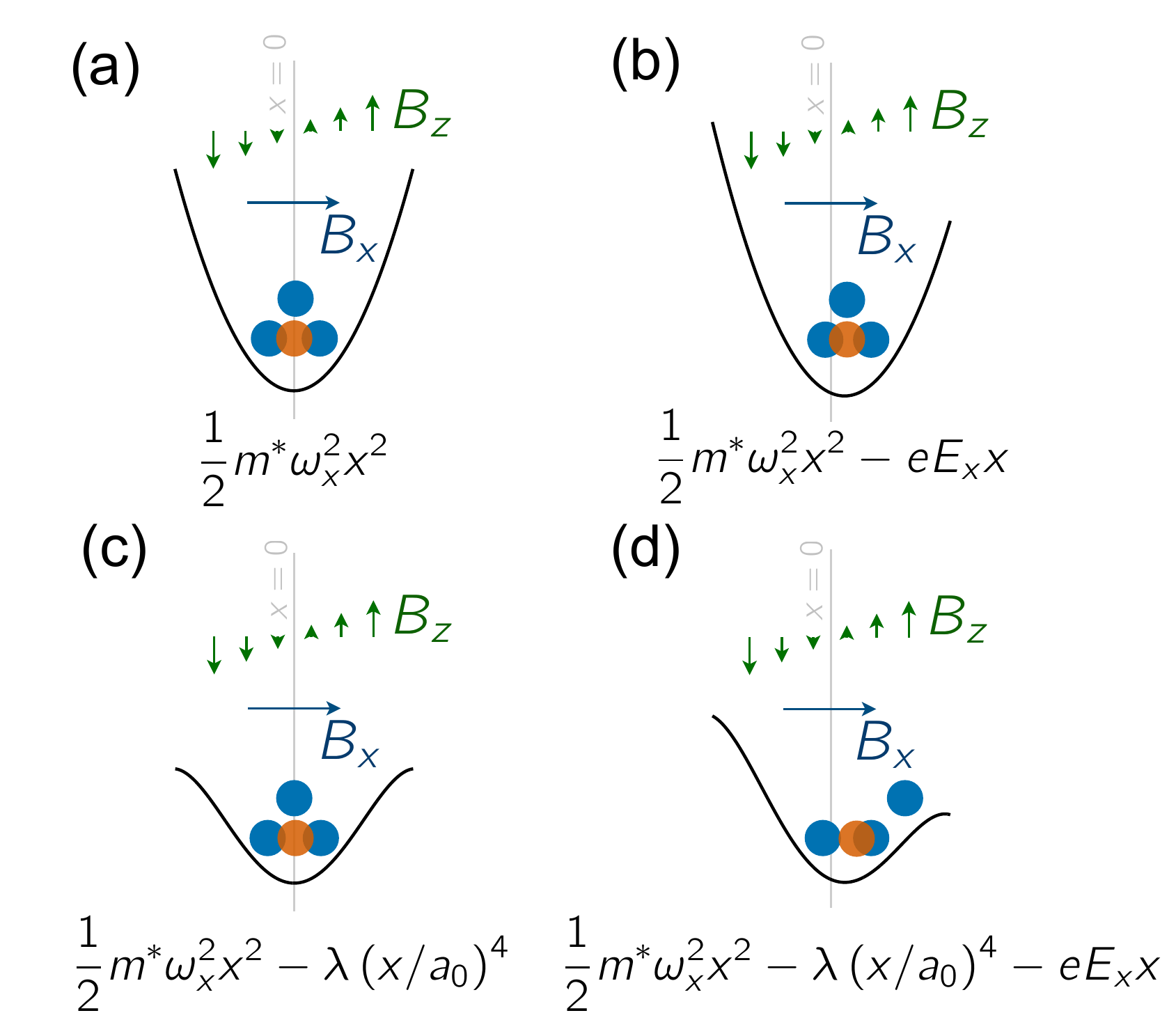}
\caption{Schematic description of setup and the effect of soft anharmonicity on 1-electron vs 3-electron motion under electric field.  Blue circles represent the electrons in a three-electron dot and a red circle represents a single electron in a dot.
(a, b) In harmonic confinement, homogeneous electric field oscillations shift all three electrons rigidly and do not result in increased spin response with respect to a single electron, in the presence of homogeneous magnetic field $B_x=B_0$ and the slanting magnetic field $B_z=bx$.
(c, d) Soft anharmonic confinement profile $\lambda<0$ leads to relative motion of electrons in the three-electron dot. Two of the electrons at the bottom form a singlet-like pair resulting in larger spin displacement due to $E_x$. In the presence of $B_x$ and $B_z$, this leads to faster Rabi oscillations with respect to a single electron.
}
\label{fig:anhCartoon}
\end{figure}

Our main results can be summarized as follows:
The strength and shape of the electrostatic confinement are both critical in multielectron EDSR. While the confinement needs to be strong or anisotropic enough to have $S=1/2$ ground state of a three-electron dot, the shape of the confinement determines the response of the electrons' spin to the driving electric field. Spin response is maximized when the core electrons are strongly confined while the valence electron is weakly confined.  Anharmonicity plays an important role here. As a time-dependent but spatially homogeneous electric field only shifts the harmonic potential without affecting its symmetry or curvature, it does not excite the internal degrees of freedom and the electrons' spin center remains mostly aligned with the center of the shifting parabolic potential profile, regardless of the electron number. This is schematically shown in Fig.\ \ref{fig:anhCartoon}(a-b). On the other hand, when the potential is anharmonic, the driving field both shifts and excites internal excitations of the electronic configuration. If the anharmonicity is soft, this results in stronger valence electron response relative to the core electrons, as schematically depicted in Fig.\ \ref{fig:anhCartoon}(c-d), leading to a higher Rabi frequency.  The steps at the quantum well interface are also sources of anharmonicity, and single steps can lead to improvement in EDSR performance for three electrons by an order of magnitude.

\begin{figure}
\includegraphics[width=3.4 in]{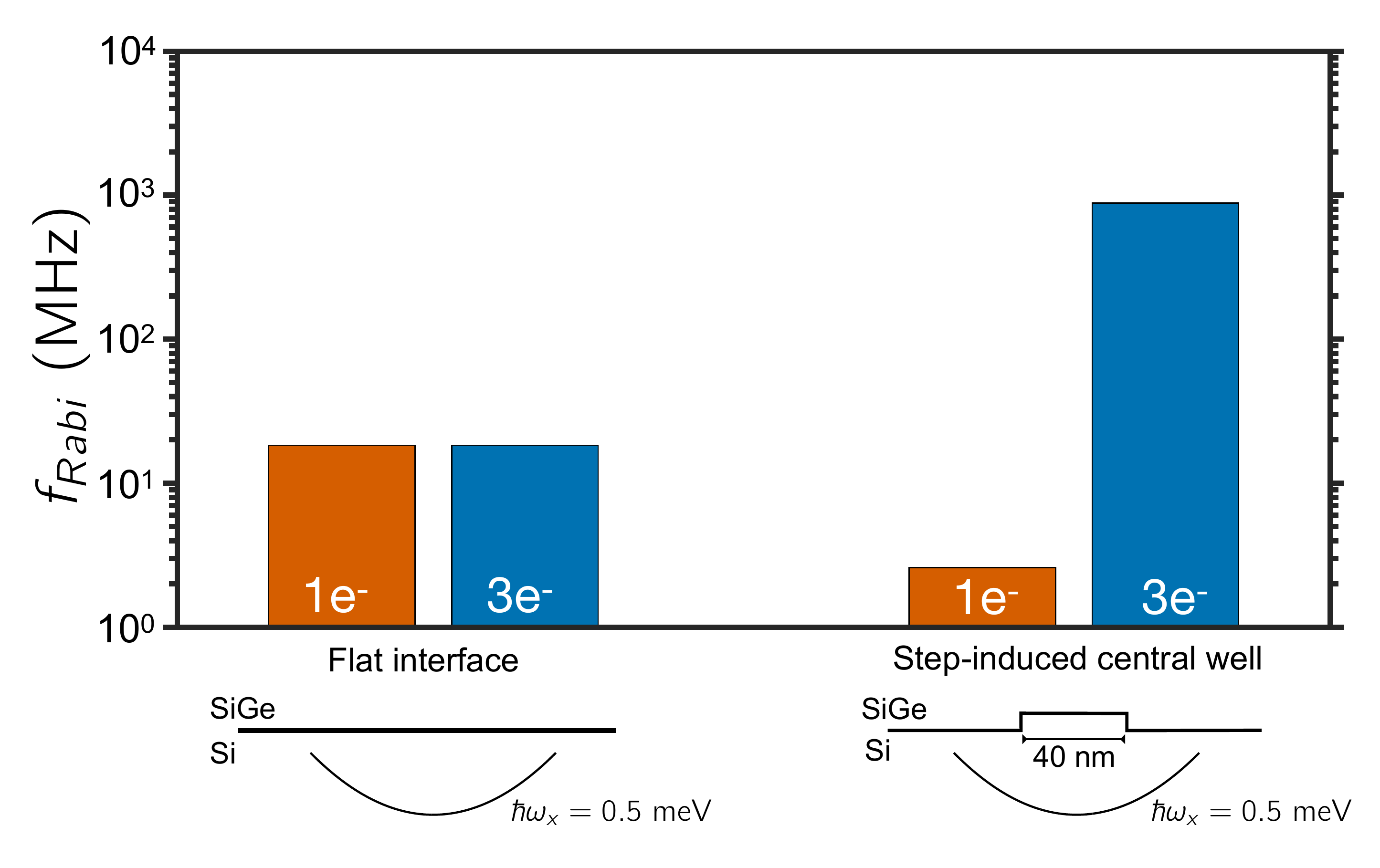}
\caption{Effect of a strong center confinement in EDSR performance.  In a harmonic confinement with flat interface, one- and three-electron performances are similar. With the addition of the center confinement induced by the steps at the interface, three-electron performance significantly increases, whereas single-electron performance drops.
}
\label{fig:summary}
\end{figure}

The multielectron spin response can be further improved, if the difference in the confinement strength of the effective core \textit{vs.}\  valence electron is increased. This can be achieved with sharp changes in the electrostatic potential.
In such a scenario, the range of the net spin motion is largely determined by the valence electron, which effectively experiences a double-dot-like potential due to the repulsion of the stable core electrons at the center of the dot, similar to the flopping mode qubit~\cite{Croot:2020p012006}. It is difficult to create such confinement solely with gates, since the potential quickly becomes smoothened away from the gates. However, interface steps can provide the desired abrupt changes in the potential. Figure~\ref{fig:summary} shows the change in EDSR Rabi frequency due to a small well created by interface steps. 
Here, the electrostatic confinement is harmonic, so the one- and three-electron responses are the same when the interface is flat. With the addition of the steps, as schematically shown in Fig.\ref{fig:summary}, the single electron Rabi frequency drops due to stronger confinement in the center of the dot, but the three-electron Rabi frequency becomes $\sim$50 times larger.
Thus, for this example the Rabi frequency can be enhanced greatly by increasing the number of electrons in the dot.

The paper is organized as follows: Sec.~\ref{Sec:Dot} describes the quantum dots in SiGe/Si/SiGe and our computational approach, followed by Sec.~\ref{Sec:EDSR} that discusses the EDSR physics and incorporation of magnetic fields in our calculations.  Later we discuss our results starting with the role of small anharmonicity in Sec.~\ref{Sec:anharmonic}. Section~\ref{Sec:EH} discusses how the EDSR performance can further be improved based on designing lateral confinement potentials in theory, and discusses prospects for achieving such potentials in realistic devices. The role of interface steps is discussed in Sec.~\ref{Sec:interface}. In Sec.~\ref{Sec:conclusion}, we give our conclusion.

\section{Electrically-driven spin resonance (EDSR) of electrons in $\mathbf{SiGe/Si/SiGe}$ quantum dots}
In this section we describe our model and computational approach, and physical quantities that are relevant to the discussion of results that follows.  Our modeling and computational approach can be divided into two main steps: (i) description of a quantum dot where any magnetic field is purely in-plane and does not affect the spatial dependence of the electronic wavefunctions, and (ii) perturbative treatment of the additional magnetic fields used to drive the EDSR to obtain an estimate of $f_{Rabi}$.

\subsection{Hamiltonian of undriven system}
\label{Sec:Dot}

We consider electrostatically defined quantum dots that are quasi-two-dimensional due to the strong quantum well confinement. We write the spatial Hamiltonian that describes a single electron in this system as
\begin{multline}
H_0=H_K^{EM}(x)+H_K^{EM}(y)+H_K^{TB}(z) \\ +H_{QW}(\bm{r})+H_{E}(\bm{r}),
\end{multline}
where $\hat z$ defines the crystallographic [001] axis and the growth direction of the heterostructure. The kinetic energy in the $x$-$y$ plane is described within the effective-mass approximation, $H_K^{EM}(x)+H_K^{EM}(y) = -(\hbar^2/2 m_t)(\partial^2 / \partial x^2+\partial^2 / \partial y^2)$, where the transverse (in-plane) effective mass $m_t=0.19 \ m_0$, with $m_0$ the electron mass, and implemented via a 6th order finite-difference approximation on a rectangular grid~\cite{Anderson:2022p065123}. We find 2.5 nm in-plane grid spacing sufficient for our purposes. To treat the two-fold degenerate conduction band valleys with minima at $\bm{k}_0=\pm0.82 (2\pi/a) \bm{\hat z}$ in the reciprocal lattice, where $a=5.43$ $\text{\normalfont\AA}$ is the Si cubic lattice constant, we use a two-band TB model with nearest- and next-nearest hopping terms in the vertical direction $z$, so that $H_K^{TB}=\sum_{i_x,i_y,i_z} t_1 \ket{i_x,i_y,i_z+1} \bra{i_x,i_y,i_z}+ t_2 \ket{i_x,i_y,i_z+2} \bra{i_x,i_y,i_z} + \text{h.c.}$, where $i_x$, $i_y$ and $i_z$ are the grid indices for $x$, $y$ and $z$ axes, respectively. In $z$, the atomic sites are separated by $a/4$ and the hopping parameters $t_1=0.68$~eV and $t_2=0.61$~eV are chosen to yield a longitudinal (out-of-plane) effective mass of Si of $m_l=0.916 \ m_0$ and the correct position of the conduction band minima, $\bm{k}_0$~\cite{Boykin:2004p115,Boykin:2004p165325,Friesen:2010p115324,Abadillo-Uriel:2018p165438,Ercan:2021p235302}.

The quantum well confinement potential arising from the difference in composition between Si and SiGe is given by $H_{QW}=V_{QW} (\Theta(z_b(x,y)-z)+\Theta(z-z_t(x,y))$, with the typical band offset $V_{QW}=150$ meV. The Heaviside step function $\Theta(z)$ describes the sharp bottom and top interfaces located at $z_b$ and $z_t$, respectively, with the dependence of $z_b$ and $z_t$ on $x$ and $y$ encoding the interface disorder. Finally, the electrostatic potential of the gates is given by $H_{E}(\bm{r})=V_L(x,y)+V_V(z)$. Here, $V_V(z)=-eE_zz$ is associated with the vertical electric field $E_z$ that pulls electrons towards the top interface, and $V_L(x,y)$ is the lateral confinement potential. In this paper, we will focus on the effects of different $H_{E}$ and $z_t$.

For three electrons, one must also account for the mutual Coulomb repulsion between the electrons. Therefore, the Hamiltonian that describes three electrons in this system is given by
\begin{equation}
H_0^{(3e)}=\sum_{i=1}^{3} H_0(\bm{r}_i) +\sum_{i=1}^{3} \sum_{j>i}^{3} \frac{e^2}{4\pi\epsilon_0 \epsilon_r}
\frac{1}{|\bm{r}_i-\bm{r}_j|}.
\end{equation}
We use the FCI method to solve the interacting electron problem~\cite{Ercan:2021p235302,Anderson:2022p065123}.
\subsection{EDSR}
\label{Sec:EDSR}
We now include magnetic fields to compute the Rabi frequency.
For a single electron the
Hamiltonian is given by:
\begin{equation}
H=H_{0}+H_{B_x}+H_{B_z}+H_t,
\end{equation}
where $H_{B_x}=\frac{1}{2}g\mu_B B_0 \sigma_x$, 
$H_{B_z}=\frac{1}{2}g\mu_B bx \sigma_z$, 
$H_t=e E(t)x$, 
$g=2$ is the g-factor, $\mu_B$ is the Bohr magneton, $B_0$ is the magnitude of the in-plane magnetic field, which is spatially uniform and temporally constant, $b$ is the coefficient describing the variation along $x$ of the slanting magnetic field pointing along $z$, $\sigma_{i}$ are the Pauli spin matrices, $E(t)=E_0 \sin{\omega_d t}$, $-e$ is the electron charge, $E_0$ is the amplitude of the AC electric field in $x$.  

When the electrostatic confinement in $x$ is harmonic, the resonant Rabi frequency is given by
%t_{\pi}\approx\frac{2\pi \hbar}{g\mu_Bb \frac{eE_0}{m_t \omega_x^2}} \left ( 1-\frac{\omega_d^2}{ \omega_x^2}\right ),
\begin{equation}
f_{Rabi} \approx \frac{g\mu_Bb |x_0|}{4\pi \hbar} \frac{1}{1-\omega_d^2 /  \omega_x^2}
\label{eq:f_rabi}
\end{equation}
where $x_0=-eE_0 / m_t \omega_x^2$ and $\hbar \omega_d=\varepsilon_Z$ is the Zeeman energy~\cite{Tokura:2006p047202}. 
In the limit $\omega_d^2 /  \omega_x^2 \rightarrow 0$
Eq.~\ref{eq:f_rabi} is the ESR Rabi frequency of an electron that is in a spatially stationary state and subject to an oscillating magnetic field of amplitude $b |x_0|$. This is the adiabatic term that accounts for the rigid motion of the electron in oscillating electric field for which the electron center of mass is $|x_0|$. The factor 
$1/(1-\omega_d^2 /\omega_x^2)$ is the nonadiabatic contribution, and is typically close to 1.  We provide a derivation of Eq.~\ref{eq:f_rabi} by switching to a reference frame moving with the dot in Appendix~\ref{sec:appEDSR1e} to make this distinction clear.  It is also possible to arrive at this expression by treating the slanting magnetic field term perturbatively, as in Ref.~\cite{Tokura:2006p047202}. 

For the three-electron case the EDSR Hamiltonian is
\begin{equation}
H^{(3e)}=H_0^{(3e)} + \sum_{i=1}^{3} \left ( H_{B_x}(i) + H_{B_z}(i)+H_t(i) \right ).
\end{equation}
Here, we generalize the approach of Ref.~\cite{Tokura:2006p047202} to three interacting electrons and implement perturbation theory up to the 2nd order in $b$ and estimate $f_{Rabi}$ from the coupling strength between the qubit states due to the driving term.  A demonstration of the generalization to three electrons is given in Appendix \ref{sec:appEDSR3e} \footnote{Appendix \ref{sec:appEDSR3e} presents a simplified calculation, where the perturbation theory is only in the first order in $b$. }.  

As in the single-electron case, $f_{Rabi}$ can also be estimated in the adiabatic approximation for multiple electrons, using the \textit{spin center} of the ground state. While the center of mass (or charge) coincides with the spin center for a single electron, in the multielectron case the charge center and the spin center can be different. The spin center is given by
\begin{equation}
x_0^S=\int x \rho_S(x,y,z) \diff x  \diff y  \diff z,
\end{equation}
where 
\begin{equation}
\rho_S = \rho_S^+ - \rho_S^-,
\end{equation}
is the spin density,
\begin{multline}
\rho_S^{\pm}(\bm{r}_1)=3 \int \Psi_0^* \Psi_0(\bm{r}_1,\alpha_1,\bm{r}_2,\alpha_2,\bm{r}_3,\alpha_3) \\ \times \delta(\alpha_1\mp1)  \diff \bm{r}_1 \diff \alpha_1  \diff \bm{r}_2 \diff \alpha_2  \diff \bm{r}_3 \diff \alpha_3,
\end{multline}
$\alpha_i$ is the $x$ projection of the $i$th electron's spin and $\Psi_0$ is the ground state wave function. We define the \textit{spin response} to be the displacement of the ground state spin center at the maxima of the driving electric field $\bar{x}_0^S = x_0^S(E=\pm E_0)$. With this definition, the Rabi frequency is estimated to be
\begin{equation}
f_{Rabi}^{(ad.)} =  \frac{g\mu_Bb |\bar{x}_0^S|}{4\pi \hbar}.
\end{equation}
This expression condenses the three-electron physics into a single parameter and provides a good estimation for the cases where the nonadiabatic effects are small, as it will be shown in Sec.~\ref{Sec:anharmonic}. However, we remind the reader that to calculate this parameter, the three-electron wave functions need to be accurately calculated  first.  Therefore, it does not dispense the need for the FCI analysis.

In most cases,  we present both the adiabatic and the nonadiabatic results. While the results including the nonadiabatic effects are expected to be more accurate, adiabatic results provide more intuitive pictures.  In all calculations, we fix the driving electric field amplitude $E_0$ at 10 $\mu$V$/\bar{a}$, where $\bar{a}=\sqrt{\hbar/m_t \bar{\omega}}$ with $\hbar\bar{\omega}=2$ meV,  the magnetic field gradient $b$ at 1.16 T$/\mu$m, and the vertical electric field $E_z$ at 6 MV/m.

\section{Results}
This section presents our results on confinement and disorder effects on multielectron EDSR.
We first describe harmonic and slightly anharmonic confinements and discuss the basic mechanism that leads to improved performance in EDSR experiments of multielectron dots compared to single-electron dots. We then extend the discussion to highly anisotropic confinements. In these subsections, we remain in the single-band approximation that is valid for a flat interface, and then we present the results for different interface conditions with the valley physics taken into account. Also, in Appendix~\ref{Sec:DoubleDot}, we discuss the our results' relation to the flopping mode EDSR in double dots.

\subsection{Nearly harmonic dots}
\label{Sec:anharmonic}
Here, we consider the case where the lateral confinement is close to harmonic; $V_L(x,y)=m_t\omega_x^2x^2/2+m_t\omega_y^2y^2/2+\lambda(x/a_0)^4$, where $a_0=\sqrt{\hbar/m_t \omega_x}$ is the harmonic dot radius in $x$, and $\lambda$ is a small anharmonicity parameter. 

For a qubit implementation, it is desirable that the ground state of the multi-spin system has total spin $S=1/2$, so that qubit transitions can be resonantly driven without causing significant leakage into the states outside of the qubit subspace. Whether this is the case or not is determined by the confinement. Therefore, it is important to draw a quantitative picture for this relationship.

\begin{figure}
\includegraphics[width=3.2 in]{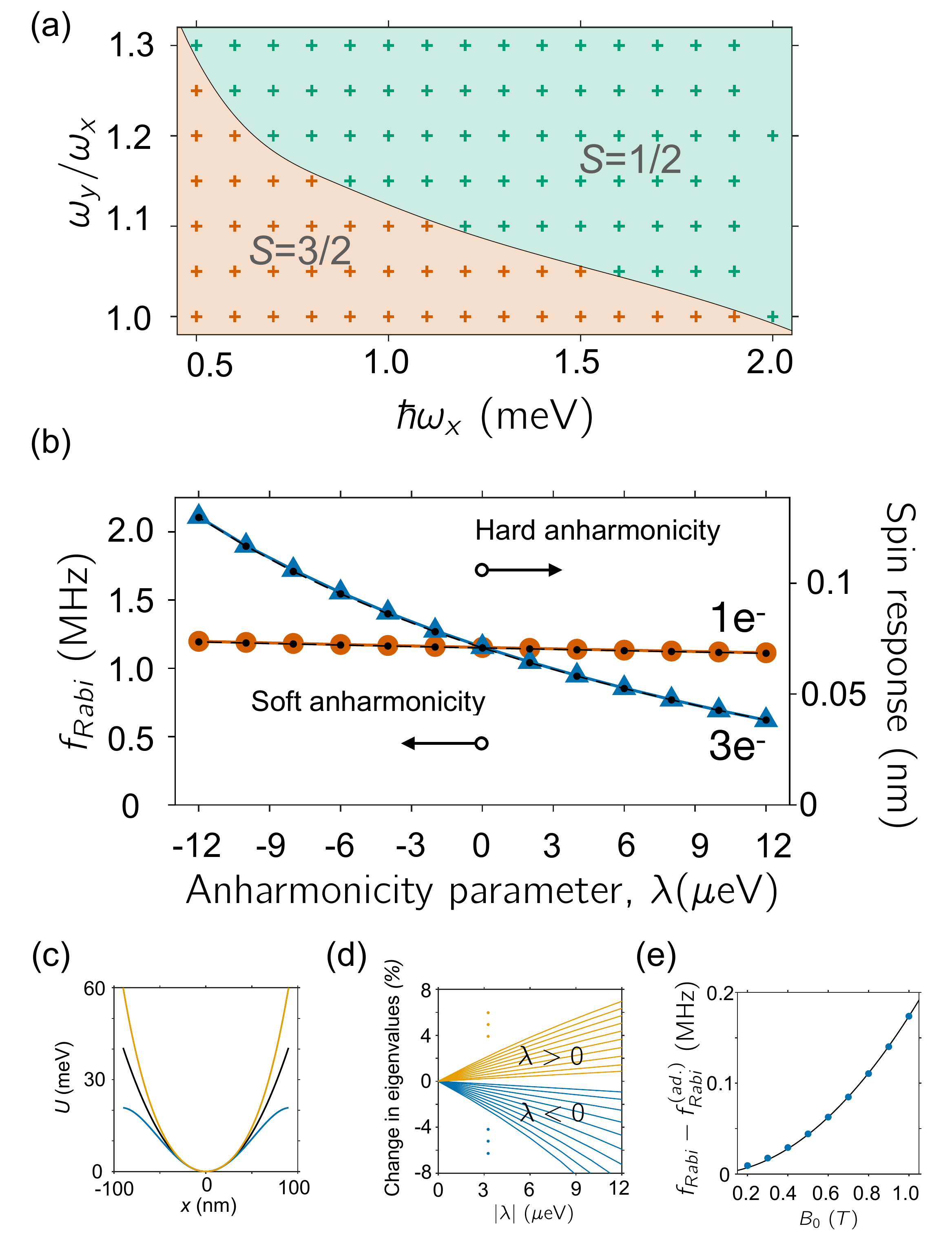}
\caption{Confinement effects on three-electron EDSR. (a) Total spin of the ground state as a function of the confinement strength and anisotropy. (b) Rabi frequency as a function of the anharmonicity parameter $\lambda$ for a single- (red circles) and three-electron (blue triangles) dot, for $\hbar\omega_x=2$ meV, $\hbar\omega_y=2.2$ meV and $B_0=0.3$ T.  The black dots show the results in adiabatic approximation, and therefore indicate that the nonadiabatic effects are small. When compared to the harmonic case, hard anharmonicity ($\lambda>0$) lowers $f_{Rabi}$ while soft anharmonicity ($\lambda<0$) increases it. (c) Confinement profiles in $x$ for $\lambda=0$ (black), $\lambda=12 \ \mu$eV (yellow), and $\lambda=-12 \ \mu$eV (blue).  (d) Percentage change in the first ten single electron energy eigenvalues going from $\lambda=0$ to $\lambda=\pm12 \ \mu$eV.  (e)The slight underestimation of $f_{Rabi}$ in the adiabatic approximation for the three-electron case at $\lambda=-12 \ \mu$eV, as a function of the in-plane magnetic field $B_0$ (blue dots). As in Eq.~\ref{eq:f_rabi}, this difference is proportional to $B_0^2$. 
The black line shows a fit of a parabola centered at $(0,0)$.
}
\label{fig:ConfData}
\end{figure}

For three electrons in an isotropic and harmonic potential, the ground state transitions  from  polarized ($S=3/2$) to unpolarized ($S=1/2$) going from weak to strong confinement. This is understood when the symmetry of the problem that only allows energy eigenstates with spin and angular momentum quantum numbers $(S,L)=(3/2,3k)$ or $(S,L)=(1/2,3k+1)$, where $k$ is an integer, is considered~\cite{Ruan:1995p7942}. As the confinement gets stronger, the energies of the states with higher orbital angular momentum grow slower than the ones with lower orbital angular value, due to the Coulomb repulsion, leading to a $(3/2,0)\rightarrow (1/2,1)$ transition in the ground state. Figure~\ref{fig:ConfData}(a) shows the total spin of the ground state of the three-electron dot at $\lambda=0$ as a function of the $x$ confinement energy $\hbar \omega_x$ and the anisotropy $\omega_y/\omega_x$. The transition takes place at $\hbar\omega_x=\hbar\omega_y \approx 1.9$ meV for the isotropic case, and at $\hbar\omega_x \lessapprox 1.9$ meV for $\omega_y > \omega_x$.

We consider a slightly anisotropic confinement, with $\hbar\omega_x$ at 2 meV and $\hbar\omega_y$ at 2.2 meV and calculate $f_{Rabi}$ as a function of $\lambda$. Figure~\ref{fig:ConfData}(b) shows the results for one and three electrons, as well as soft ($\lambda<0$) and hard ($\lambda>0$) anharmonicity. (To provide a visual reference, potential profiles for $\lambda=0$ and $\lambda=\pm12$ $\mu$eV are shown in Fig.~\ref{fig:ConfData}(c), and the changes in first ten single-electron energy eigenvalues as a function of $\lambda$ are plotted in Fig.~\ref{fig:ConfData}(d).) Soft anharmonicity allows faster Rabi oscillations as the center of the dot confines the electrons more strongly. This makes the core electrons less sensitive to the driving electric field while the valence electron, being subject to the repulsion of the core electrons and experiencing a weaker confinement, becomes more sensitive to it. Hard anharmonicity has the opposite effect and it makes the valence electron less sensitive to the driving electric field. The blue line with the triangular markers in Fig.~\ref{fig:ConfData}(b) show the resulting difference in $f_{Rabi}$ in these two regimes for three electrons.
On the other hand, the effect of anharmonicity on the Rabi frequency is much-less pronounced for a single electron, as demonstrated by the red line with circular markers.  This is because a single electron is only affected by the change in the confinement around the dot center, since the spatial extent of the charge density is smaller than in the three-electron case. It is also important to note that when the confinement is harmonic ($\lambda=0$) the response is essentially the same for one and three electrons. This is because the spatially homogeneous driving field only shifts the harmonic dot and in the adiabatic limit this translates into shifting the electron cloud rigidly. The center of spin remains aligned with the center of the dot in either case, resulting in the same response. 

While our main approach for calculating $f_{Rabi}$ is based on perturbation theory, as in Ref.~\cite{Tokura:2006p047202}, it is also useful to consider adiabatic motion of the electron due to the driving electric field and estimate $f_{Rabi}$ based on the range of spin motion in the slanting magnetic field.  In Fig.~\ref{fig:ConfData}(b), the black dots are the values based on this approach. Based on Eq.~\ref{eq:f_rabi}, it is expected that the error made in this process scales as $B_0^2$ for $f_{Rabi}^{-1}$.  We check this by calculating the error as a function of $B_0$. Figure \ref{fig:ConfData}(e) shows this data (blue dots), and a fitted parabola that goes through $(0,0)$ (black curve).

\subsection{Terraced potential}
\label{Sec:EH}
\begin{figure*}
\includegraphics[width=7 in]{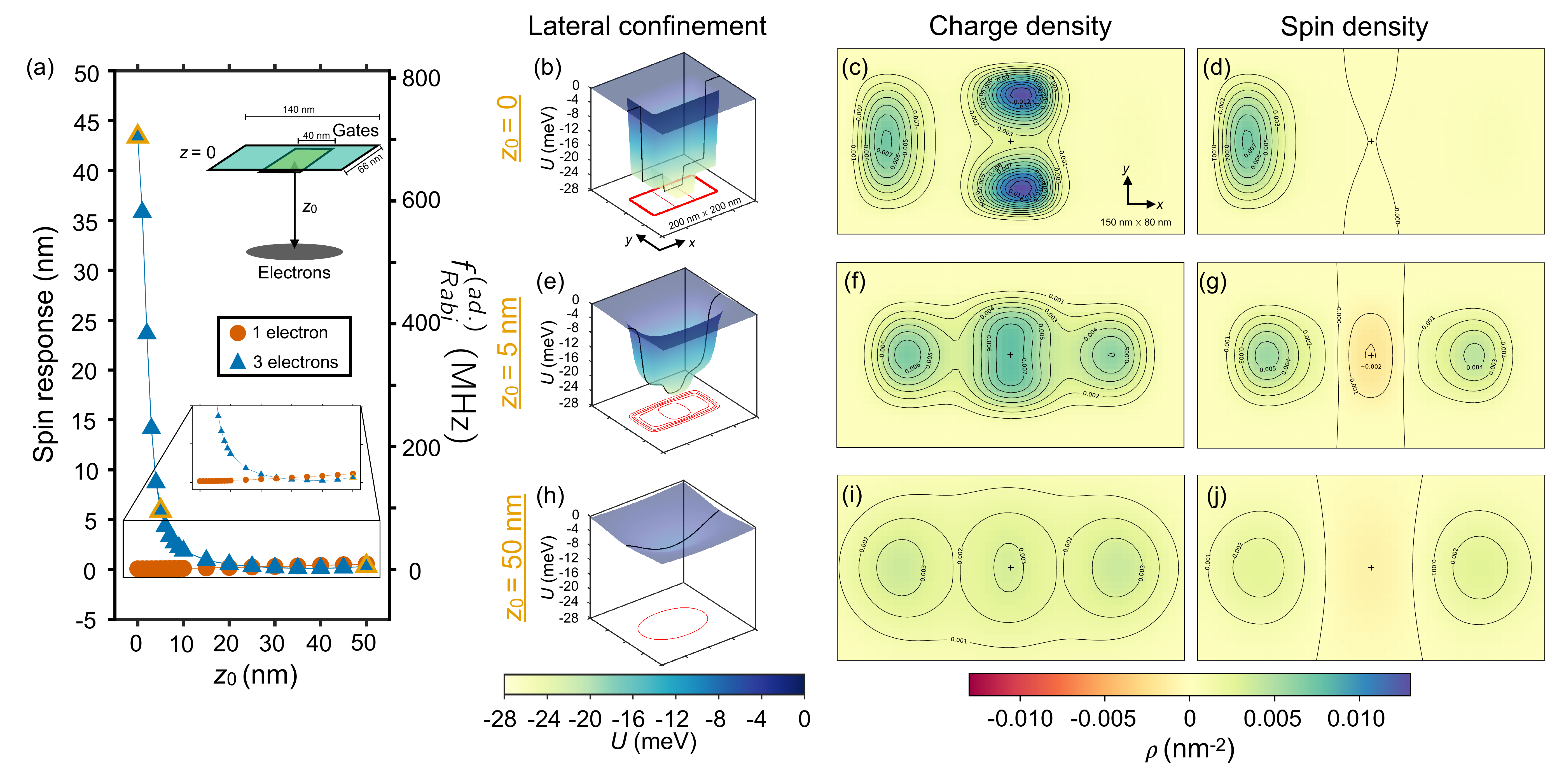}
\caption{Spin response in a confinement potential induced by two overlapping rectangular gates of different sizes to the electric field $E_0 \bm{\hat x}$. (a) Spin response as a function of the depth of the quantum dot, $z_0$. Triangle markers with the yellow edges represent the cases corresponding to (b-d)($z_0=0$), (e-g)($z_0=5$ nm), and (h-j)($z_0=50$ nm).  $f_{Rabi}$ is estimated in adiabatic approximation. Upper inset: Schematic of the geometry. Lower inset:While the three-electron response generally decreases with increasing $z_0$, the single-electron response has the opposite behavior. (b, e, h) Lateral confinement potentials in the absence of the driving electric field.  Both 3D (blue-yellow) and contour (red) plots of the potential are shown. Dark curves indicate cuts along the $x$ axis. (c, f, i) Ground state charge densities in the presence of the driving electric field. (d, g, j) $S=1/2$, $S_x=1/2$ ground state spin densities in the presence of the driving electric field. In the density plots, the black + marker denotes the origin.
}
\label{fig:EH}
\end{figure*}

In Sec.~\ref{Sec:anharmonic}, we have shown that slightly (soft) anharmonic confinements lead to faster Rabi oscillations by increasing the spin response to electric field and attributed this to valence electron being less-strongly confined than the core electrons. Here, we discuss potential profiles that can further strengthen this behavior, and the practical aspects regarding their implementation. 

To achieve a three-electron configuration with a core insensitive to the electric field but with a sensitive valence electron, we consider a 5 meV deep $40\times66$ nm rectangular well inside another 21 meV deep, $140\times66$ nm one. Creating a large contrast between the potentials experienced by the core and valence electrons, this configuration yields a very large spin response, comparable to the flopping mode. Such a potential can be realized right below two overlapping rectangular gates at the surface, as illustrated in the inset to Fig.~\ref{fig:EH}(a). Potentials due to this simple gate geometry is calculated analytically, as in Ref.~\cite{Anderson:2022p065123}. We show the resulting spin response and the corresponding adiabatic estimation of $f_{Rabi}^{(ad.)}$ as a function of the distance $z_0$ between the gates and the dot electron(s) in Fig.~\ref{fig:EH}(a).  The fast decay of the three-electron response (blue line with triangular markers) is due to the fact that the potential induced below the surface rapidly becomes smoothed as its Fourier components are suppressed as $\exp(-z_0\sqrt{k_x^2+k_y^2})$, where $k_x$ and $k_y$ are the wave vector components of the in-plane potential at the boundary, as shown in Fig.~\ref{fig:EH}(b).  Being strongly confined at the center of the dot, a single electron's response (red line with circular markers) remains low but slightly increases
as the potential becomes smoother, as also shown in Fig.~\ref{fig:EH}(a).

Three cases highlighted by triangular markers with yellow edges in Fig.~\ref{fig:EH}(a) are characterized further in Fig.~\ref{fig:EH}(b-j). In the first case, shown in Fig.~\ref{fig:EH}(c), the surface potential is taken to be the confinement potential. Of course, it is not possible to form a quantum dot right at the surface but it is useful for demonstrating the effect of having sharp features in the confinement potential. The charge density of the ground state in Fig.~\ref{fig:EH}(c), indicates that the core pair of electrons are localized in the central well, while the valence electron is pushed away from the center by the driving field. The corresponding spin density in Fig.~\ref{fig:EH}(d) also indicates that the core pair is essentially a singlet (spin-0) and the net spin is concentrated on the valence electron. Being subject to the Coulomb repulsion of the core pair, the valence electron effectively experiences a double dot potential, resulting in flopping mode-like behavior. The spin response of $\sim$43 nm is nearly three orders of magnitude larger than the response of a single electron in this potential, which is $\sim$0.07 nm. 

For the cases with $z_0=5$ nm (Fig.~\ref{fig:EH}(e-g)) and $z_0=50$ nm (Fig.~\ref{fig:EH}(h-j)), which are relevant distances for SiMOS and Si/SiGe devices, respectively, the electrons transition from the isosceles triangle configuration in Fig.~\ref{fig:EH}(c), to essentially linear configurations in $+-+$ spin formation, as denoted in Fig.~\ref{fig:EH}(f,i) and Fig.~\ref{fig:EH}(g,j), resulting in less-localized spin. For $z_0=5$ nm, the spin response of $\sim$6 nm is nearly two orders of magnitude larger than the response of a single electron in the same potential, which is $\sim$0.08 nm. Because the enhancement is largest when the electrons in the quantum dot are close to the gates defining the confinement potential, SiMOS devices are better candidates than Si/SiGe structures for engineering confinement potentials to increase spin response with multielectron dots.

\subsection{Effects of interface steps}
\label{Sec:interface}

\begin{figure*}
\includegraphics[width=7 in]{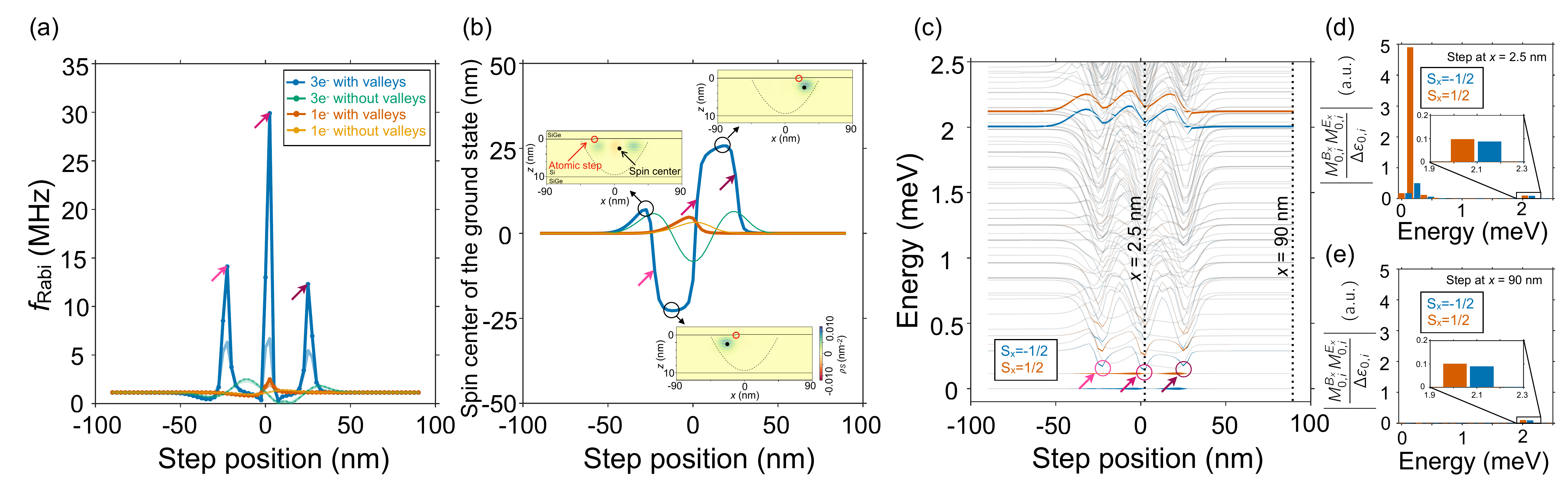}
%\caption{Effect of an atomic interface step on EDSR performance. (a) One-(red) and three-electron $t_{\pi}$ (blue) as a function of the atomic step at the top interface. Due to the strong vertical electric field of 6 MV/m, the bottom interface has a negligible effect. Black dots are the $t_{\pi}$ values estimated from the distance the spin travels with $E_x$ going from $\pm \tilde{E}_0$ to $\mp \tilde{E}_0$, in the adiabatic approximation. (b) The Si/SiGe interfaces and the three-electron ground state charge density in the $xz$ plane. The lateral confinement potential is harmonic with $\hbar\omega_x=\hbar \omega_y/2=$2 meV. (c) The spin center at $E_x=\pm \tilde{E}_0$ for one- (red) and three-electrons (blue) in the adiabatic approximation. The difference between $-\tilde{E}_0$ and $\tilde{E}_0$ values are small compared to the changes in spin centers due to the step position, and are shown on the right-hand vertical axis for one-(yellow) and three-electrons (green).
%}
\caption{Effect of an atomic interface step on EDSR performance.  (a) Rabi frequency as a function of the step position for a single electron and three electrons in parabolic confinement with $\hbar \omega_x=\hbar \omega_y/2=2$ meV and uniform in-plane field $B_0=1$ T.  The Rabi frequency of the three-electron dot increases much more rapidly than that of the single-electron dot when the step is near the dot.  The Rabi frequency also has three different high-performance spots (indicated with colored arrows) as opposed to one in the single-electron case. Rabi frequencies that are calculated without the valley physics taken into account are also shown for a single-electron (yellow) and for a three-electron dot (green), showing the important role of valley physics. The lighter colored curves show the Rabi frequencies that are calculated in adiabatic approximation for all four cases considered. (b) Spin center of the ground state as a function of the step position for the four cases considered in (a) (with the same color coding), in the absence of a driving electric field.  The location of the spin center of the three-electron dot is more sensitive to the step position than its single-electron counterpart. Including the valley degree of freedom also leads to larger responses to the step position. The derivative of these curves approximates the sensitivity to small electric field oscillations in EDSR, for fixed step positions, and are used to calculate the adiabatic approximations in (a). The three colored arrows show the three high-slope locations corresponding to the three peaks in (a).  Inset: Spin densities at the three locations that lead to the largest displacement of spin centers due to the interface step, in the $xz$ plane. (c) Three-electron eigenenergies for spin projections $S_x=\pm 1/2$ as a function of the step position. Thickness of the lines are proportional to the corresponding states' contribution to the coupling between the qubit states, $|M^{B_x}_{0,i} M^{E_x}_{0,i}|$, where $M^{B_x}_{0,i} \propto \bra{\Psi_0^{+(-)}}\sum_{j=1}^3 x_i \sigma_i\ket{\Psi_i^{-(+)}}$ and $M^{E_x}_{0,i} \propto \bra{\Psi_0^{+(-)}}\sum_{j=1}^3 x_i\ket{\Psi_i^{+(-)}}$. When the step is far from the electrons the only significant contribution comes from the states around 2 meV, which corresponds to the center of mass excitations. In these regions the electron motion is adiabatic.  The three colored circles show the regions where there is a low-lying excited level with significant coupling to the nearby qubit level.  These regions correspond to the three peak points in (a) marked by colored arrows, where non-adiabatic effects become important. (d-e) Contribution of the unperturbed energy eigenstates to the coupling of the qubit states weighted by the energy separation $\Delta \varepsilon_{0,i} = \varepsilon_i^{+(-)} - \varepsilon_0^{-(+)}$, (d) when the step is near the dot center, (e) when the step is far from the dot center.}
\label{fig:dis1}
\end{figure*}

Interface disorder is unavoidable in Si/SiGe interfaces and it strongly affects the qubit behavior. Here, we first consider a single atomic interface step (height of $a/4$, where $a$ is the silicon cubic lattice parameter) at various locations as a simple model for interface disorder and analyze its influence on EDSR performance. We then discuss a specific interface profile that leads to a flopping-mode-like behavior as in Sec.~\ref{Sec:EH}.

For simplicity, we assume that the interface profile is independent of $y$. The lateral confinement is harmonic with $\hbar\omega_x=\hbar \omega_y/2=$2 meV, and centered at $x=0$. Relatively high anisotropy in confinement is chosen to ensure that the qubit states remain as the lowest two energy eigenstates in the presence of $B_x=1$ T, which can also be achieved with lower anisotropy if the overall confinement is chosen to be stronger or $B_x$ is chosen to be lower. As seen in Fig.~\ref{fig:dis1}(a), $f_{Rabi}$ in the three-electron case (blue) significantly increases at three step locations near the dot, much more so than the single-electron case (red), which peaks at a single step location. To demonstrate the important role that the valley-orbit coupling (VOC) plays in enhancing $f_{Rabi}$~\cite{Huang:2017p75403}, we also show $f_{Rabi}$ values calculated without the valley physics being taken into account for single- (yellow) and three-electron (green) cases~\footnote{Valley physics can be turned off by simply replacing $H_K^{TB}$ with -($\hbar^2/2m_l)\partial^2/\partial z^2$}.
Including the valley physics results in a large increases in the Rabi frequency for both the one- and three-electron cases.

There are two factors leading to the enhancement of $f_{Rabi}$ in the presence of an interface step: (i) enhanced adiabatic motion of the spin due to VOC, (ii) non-adiabatic effects due to the suppression of excitation energies. The lightly-colored companions of the four curves in Fig.~\ref{fig:dis1}(a) are the corresponding adiabatic estimations of the Rabi frequency, $f_{Rabi}^{(ad.)}$, and they show that the non-adiabatic effects are especially important in the three-electron case. 

To understand the behavior of $f_{Rabi}^{(ad.)}$, we plot the spin centers as a function of the step position in Fig.~\ref{fig:dis1}(b) at zero driving electric field. Since the driving field essentially shifts the dot, the spin response is also estimated by the slope of these curves yielding the adiabatic estimations in Fig.~\ref{fig:dis1}(a). Here, the behavior of the single electron (red) is easier to understand. As the step position approaches the dot from the left, the electron moves to the right to lower its potential energy, until the displacement becomes so large that its potential energy starts to increase due to the harmonic confinement. As the step position moves further to right, the electron returns to its initial position. The three-electron behavior (blue) is expectedly more complicated as in addition to the lateral and step induced confinements, it is also determined by the mutual Coulomb repulsion and resulting relative motion of the electrons. In the inset of Fig.~\ref{fig:dis1}(b), we show the spin densities in the $xz$ plane at the three extrema of the spin center. When the step is at $x=-22.5$ nm, the electrons are essentially in $+-+$ spin configuration and the shift in the spin center is mainly due to the overall shift of the electrons. When the the step is at $x=-7.5$ nm, the two right-most electrons form a singlet-like pair with spin-0 and the net spin is localized on the left-most electron causing the spin center to shift to the left. Finally, when the spin center is at $x=15$ nm, the singlet-like pair forms on the left, leading to the spin center shift to right. The colored arrows in Fig.~\ref{fig:dis1}(b) show the corresponding high $f_{Rabi}$ locations, similarly denoted in Fig.~\ref{fig:dis1}(a), and coincide with the high slope locations indicating high sensitivity to the change in dot position in $x$. We also note that inclusion of the valley physics leads to significant quantitative changes both in single- and three-electron behavior.

Nonadiabatic effects become important in determining $f_{Rabi}$ when there are excited states close to the qubit levels. In Fig.~\ref{fig:dis1}(c), we show the three-electron energy levels (in the absence of the slanting magnetic field and driving electric field) as a function of the step position.  The thickness of the levels indicates the combined strength of the coupling matrix elements that contribute to $f_{Rabi}$, as explained in the caption, with blue (red) color labeling the spin projection $S_x=-1/2$ ($S_x=1/2$). When the step is positioned away from the dot the coupling strength of the qubit states mainly couples via the center-of-mass excited states at $\sim$2 meV (see Appendix~\ref{sec:appEDSR3e}). Therefore, in these regions the electron motion is adiabatic. At the three high-$f_{Rabi}$ positions that are marked by colored arrows (corresponding to the ones in Fig.~\ref{fig:dis1}(a,c)), an excited level with significant combined coupling becomes close to the higher-energy qubit level. The overall contributions of the coupling matrix elements and the energy separations for the step positions $x=2.5$ nm and $x=90$ nm are shown in Fig.~\ref{fig:dis1}(d, e).

\begin{figure}
\includegraphics[width=3.4 in]{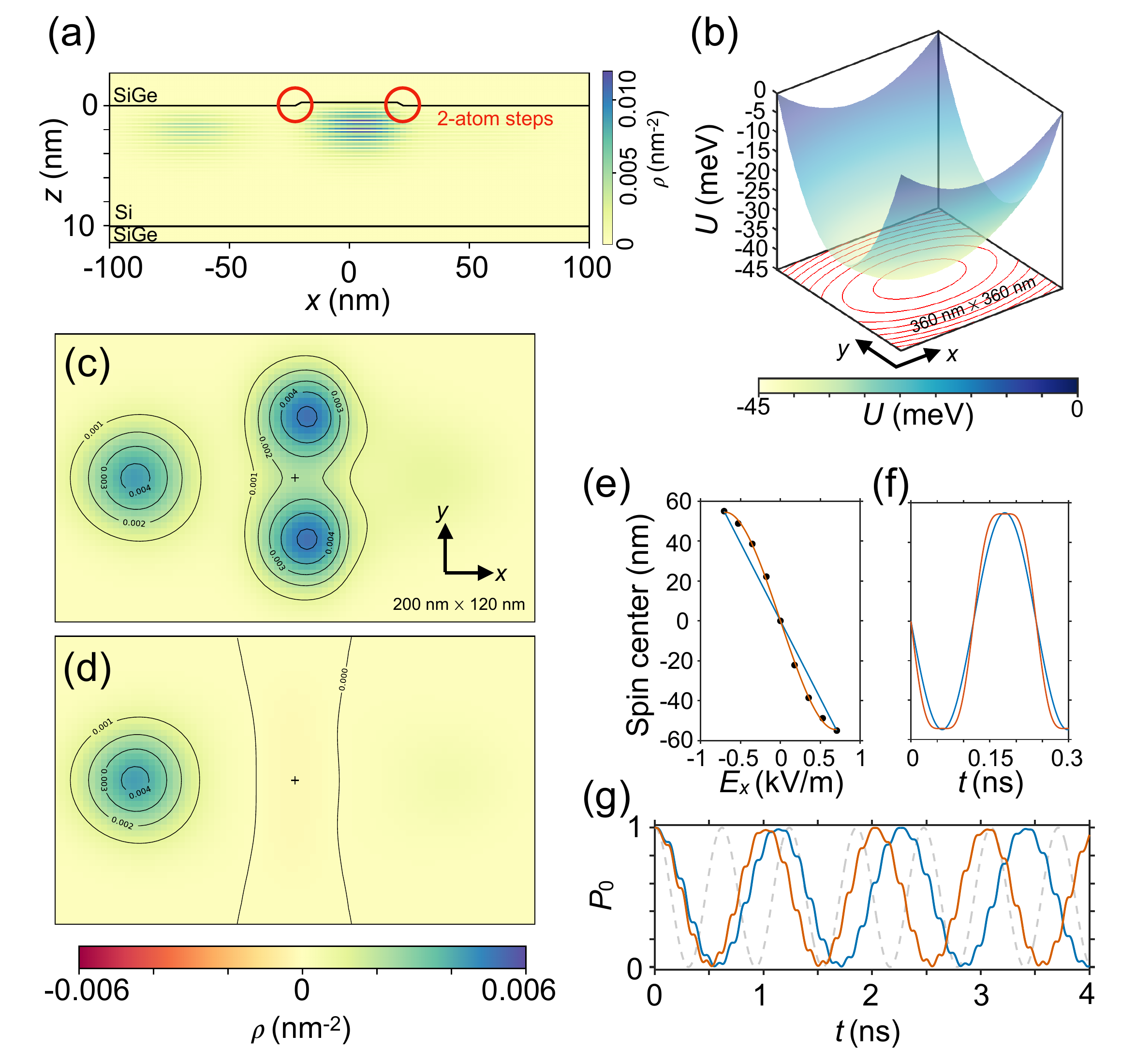}
\caption{Three-electron flopping mode in a single dot with interface steps.  (a) Two-atom interface steps and the three-electron charge density at electric field $E_0 \bm{\hat x}$. 40 nm-wide dip at the interface creates a central well that confines the core electrons together with the strong vertical electric field of 6 MV/m.  (b) The lateral confinement potential is harmonic with $\hbar\omega_x=0.5$ meV and $\hbar\omega_y=0.9$ meV. (c) Ground state charge density at electric field $E_0 \bm{\hat x}$. (d) Ground state spin density at electric field $E_0 \bm{\hat x}$. (e) Spin center as a function of the in-plane electric field $E_x$ ranging from  $-E_0$ to $E_0$ in adiabatic approximation (black dots). The linear fit based only on the extrema (blue) and the third-order polynomial fit (red) are used to estimate $f_{Rabi}$. (f) Spin center as a function of time as $E_x$ varies sinusoidally, when linear response is assumed (blue), and when the nonlinearity of the response is taken into account. (g) Time evolution based on nonlinear (blue) response and its linear approximation (red) are close to each other with $f_{Rabi}^{(ad.)}\approx1$ GHz at $B_0=0.15$ T. Perturbative calculation yields $f_{Rabi}=1.6$ GHz and the corresponding $\cos^2{(\pi f_{Rabi}t)}$ oscillations are shown in the background (gray-dashed).  The difference indicates nonadiabaticity plays a role.  
}
\label{fig:dis2}
\end{figure}

Although a single atomic step at the interface can lead to $\sim$30-fold improvement in three-electron $f_{Rabi}$ (and a factor of two improvement for a single electron), it is not as efficient as the potential profile in Fig.~\ref{fig:EH}(b). However, a similar effect can be provided by two steps at the interface. Here, we consider two steps in alternating directions that are of two atoms height (height of $a/2$) separated by 40 nm, as shown in Fig.~\ref{fig:dis2}(a), together with the ground state charge density at electric field $E_0 \bm{\hat x}$. The lateral confinement is taken to be parabolic with $\hbar \omega_x=0.5$ meV and $\hbar \omega_y=0.9$ meV, as also shown in Fig.~\ref{fig:dis2}(b). As the isosceles triangle charge configuration in Fig.~\ref{fig:dis2}(c), and the localization of spin on the valence electron in Fig.~\ref{fig:dis2}(d) indicate, this is indeed similar to the case considered in Fig.~\ref{fig:EH}(b-d).

In estimating $f_{Rabi}$, in the adiabatic approximation, we also consider the effect of nonlinear response.  Figure \ref{fig:dis2}(e) shows the adiabatic motion of the spin center as a function of $E_x$ (black circles), together with a linear fut based on the two extreme values  (blue line), and a third order polynomial fit based on all data points. Resulting spin motion during the sinusoidal drive as a function of time is shown in Fig.~\ref{fig:dis2}(f). Numerical simulation of the time evolution based on the two response functions yield similar $f_{Rabi}$ values around 1 GHz, as the Rabi oscillations in Fig.~\ref{fig:dis2}(g) show.  Due to the low-lying energy levels a smaller Zeeman field, $B_0=0.15$ T, is chosen in this case so that the qubit energy levels remain the two lowest-lying levels. The fast modulations in Rabi oscillations are attributed to strong driving~\cite{Yang:2017p062321}. Perturbative calculation of $f_{Rabi}$ yields a slightly higher value. For comparison, we also include corresponding oscillations in Fig.~\ref{fig:dis2}(g) (gray-dashed line). 

Although creating interface conditions as in Fig.~\ref{fig:dis2}(a) in a predictable and feasible manner require fundamental advances in technology, it shows that a combination of gate-induced electrostatic confinement and controllable interface conditions can lead to novel design considerations for improving qubit performance. Possible construction techniques may include reactive ion etching and STM-based approaches~\cite{Huang:2017p75403}.

\section{Summary and conclusions}
\label{Sec:conclusion}

In this paper, we have studied the conditions that lead to faster Rabi oscillations with increasing electron number in electrically-driven spin resonance (EDSR) in Si/SiGe quantum dots by performing full-configuration interaction (FCI) and tight-binding calculations to accurately account for the electron-electron (e-e) interactions and the valley physics in the presence of interface steps.

The electrostatic landscape that the electrons are subject to affects the EDSR performance in two ways: (i) It determines the spin configuration of the ground state, and (ii) it determines the response of the spin to the driving electric field.  For three electrons considered in this paper, the ground state can either have a total spin of $1/2$ or $3/2$.  Spin-$1/2$ is desirable for qubit implementations and it can be achieved in stronger confinements, in general.  The spin-electric coupling depends both on the relative and the collective movement of the electrons in response to the driving electric field, and therefore is sensitive to the interplay between the electrostatic environment and e-e interaction effects.

We have focused on two factors that are important in determining the electrostatic landscape: (i) the potential induced by the gates, and (ii) steps at the heterostructure interface.  Modeling the gate-induced potentials with harmonic and slightly anharmonic (with a quartic term) functions shows that a soft anharmonic confinement is needed to benefit from increasing electron number in an EDSR experiment. While the spin response of a single- and a three-electron system in harmonic confinement is essentially the same, in the case of a soft (hard) harmonic confinement the three-electron response is higher (lower). This effect intensifies as the core electrons become more confined (therefore less responsive to the electric field) at the center and the valence electron become less confined. Since the valence electron is subject to the repulsion of the core electrons it essentially experiences a double-dot-like potential leading to high response as in the case of single-electron flopping mode qubits.

Confinements required for flopping-mode-like high responses need sharp features to host an insensitive core and a sensitive valence electron to the driving field. Such potentials can be realized at the surface of the device using overlapping gates, however,  they get smoothened quickly as the distance from the surface to the quantum dot electrons grows. SiMOS dots may partially benefit from this as the surface-dot separation is smaller ($\sim$5 nm); for Si/SiGe dots with larger separations ($\sim$50 nm) this effect reduces to slight anharmonicity.

\begin{figure}
     \centering
     \includegraphics[width=3.4 in]{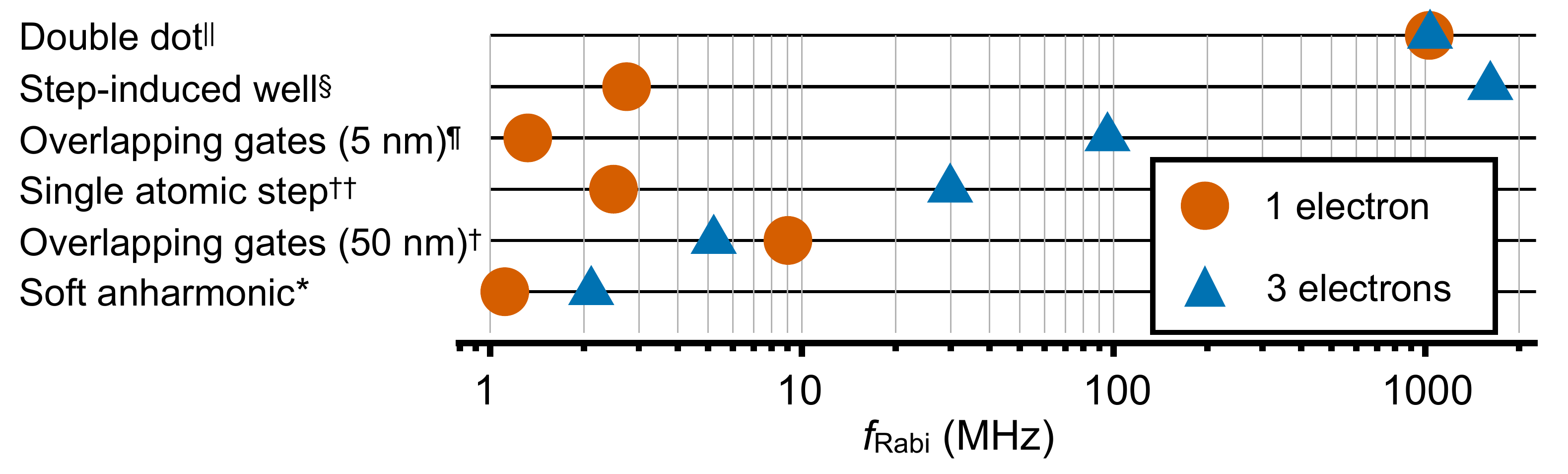}
     \caption{Summary of the results. Highest Rabi frequencies obtained under different conditions in this paper are shown: *Fig.~\ref{fig:ConfData}, $^\dagger$Fig.~\ref{fig:EH}, $^{\dagger \dagger}$Fig.~\ref{fig:dis1}, $^\P$Fig.~\ref{fig:EH}, $^\S$Fig.~\ref{fig:dis2}, $^{||}$Fig.~\ref{fig:doubleDot}.}
     \label{fig:end_summary}
 \end{figure}

Steps at the heterostructure interface are ubiquitous and can provide sharp changes in the confinement potential when combined with the gate-induced strong vertical electric field that pulls electrons towards the interface. In particular, a pair of two atomic layer deep steps in alternating directions with the correct separation can create a central mini-well that strongly confines the core electrons, but not the valence electron and yield a spin response that is about 50 times the response in the presence of a flat interface. For creation of such interface to become viable, fundamental advances in interface engineering is needed, however, this exemplifies a potential benefit that lies ahead. On the other hand, single steps that are in general in the same direction (creating tilts at the interface) are commonly found at interfaces. A single step at the interface can lead to an order of magnitude improvement in Rabi frequency when it is near the dot. The sensitive dependence of the Rabi frequency on the position of the step also highlights the kind of variation in Rabi frequencies that could manifest in experimental data given the relatively uncontrolled atomic structure of material interfaces. We summarize the best results in various conditions in Fig.~\ref{fig:end_summary}.

In conclusion, this study provides a theoretical understanding of the conditions leading to improved performance of multielectron EDSR that has been experimentally reported and discusses the opportunities and limitations for further improvements.

\begin{acknowledgments}
The authors thank Andrew Pan, Samuel Quinn,  Andrey Kiselev, Joseph Kerckhoff,  Efe \c{C}akar, Merritt Losert, Andre Saraiva, and Jos\'{e} Carlos Abadillo-Uriel for helpful discussions, and acknowledge the support from HRL Laboratories. M. \ F.\ G.\ and C.\ R.\ A.\  acknowledge a financial interest in HRL Laboratories as paid consultants on projects unrelated to this study. This research was sponsored in part by the Army Research Office (ARO), through Grant Number W911NF-17-1-0274. The views and conclusions contained in this document are those of the authors and should not be interpreted as representing the official policies, either expressed or implied, of the Army Research Office (ARO), or the U.S. Government. The U.S. Government is authorized to reproduce and distribute reprints for Government purposes notwithstanding any copyright notation herein. 
\end{acknowledgments}
\appendix
\section{Single electron EDSR in harmonic confinement}
\label{sec:appEDSR1e}
It is illuminating to arrive at the ESR frequency for a harmonic dot with a single electron, Eq.~\ref{eq:f_rabi}, by switching to the moving dot frame~\cite{Rashba:2008p195302}. Noting that a homogeneous electric field $E_0$ shifts the center of the parabolic confinement by $x_0=-eE_0/m_t\omega_x^2 $, one transforms the Schr\"{o}dinger equation $i\hbar \frac{\partial}{\partial t}\psi=H \psi$ to $i\hbar \frac{\partial}{\partial t}\tilde{\psi}=\tilde{H} \tilde{\psi}$, where $\tilde{\psi}=U \psi$, $\tilde{H}=UHU^{\dagger} +i \hbar \frac{\partial U}{\partial t}U^{\dagger}$, and $U=e^{-\frac{i}{\hbar}p_x  x_0 \sin{(\omega_d t)}}$. The Hamiltonian in this frame becomes
\begin{multline}
\tilde{H}= H_0 +H_{a}(t) +H_{B_x}+H_{f}(t) + H_{B_z}.
\end{multline}
Here, the newly emerging terms are the adiabatic coupling to the slanting magnetic field $H_{a}(t)=\frac{1}{2}g\mu_B b x_0 \sin{(\omega_d t)} \sigma_z $ and  the fictitious potential $H_{f}(t)=\frac{e E_0 \omega_0}{m_t\omega_x^2} p_x  \cos{(\omega_d)}t$.  Also, note that the term $- e^2 E(t)^2 /2m_t\omega_x^2$ can be ignored as it is an overall energy shift. The first three terms essentially describe ESR, as the adiabatic motion of an electron in slanting magnetic field is equivalent to an electron in a spatially stationary state subject to an oscillating magnetic field.  $H_f(t)$ is associated with the non-inertial nature of the oscillating frame, and together with $H_{B_z}$ it introduces the second order correction to $t_{\pi}=f_{Rabi}^{-1}/2$ in the nonadiabaticity parameter $\omega_d/\omega_x$.

To see this, consider the eigensystem of $H_0+H_{B_x}$:
\begin{equation}
(H_0+H_{B_x}) \ket{n,\pm}=\varepsilon_n^{\pm} \ket{n,\pm},
\end{equation}
where $\varepsilon_n^{\pm} = n \hbar \omega_x + \hbar \omega_x/2 + \varepsilon^{y}_0 \pm \varepsilon_Z/2$.
First order in $b$, the qubit states $\ket{Q_0}=\ket{0,-}$ and $\ket{Q_1}=\ket{0,+}$ become
\begin{subequations}
\begin{equation}
\ket{Q_0}=\ket{0,-} +\frac{1}{2}g\mu_B b \frac{\bra{1,+} x \sigma_z \ket{0,-}}{\varepsilon_0^- - \varepsilon_1^+} \ket{1,+},
\end{equation}
\begin{equation}
\ket{Q_1}=\ket{0,+}  + \frac{1}{2}g\mu_B b \frac{\bra{1,-} x \sigma_z \ket{0,+}}{\varepsilon_0^+ - \varepsilon_1^-}\ket{1,-}.
\end{equation}
\end{subequations}
Noting that $\bra{1} x \ket{0}=\sqrt{\hbar/2m_t \omega_x}$, $\bra{1} p_x \ket{0}=i\sqrt{\hbar m_t \omega_x/2}$ and $\bra{\pm}\sigma_z\ket{\mp}=1$, the qubit Hamiltonian becomes (after a $\pi/2$ rotation around $z$ for convenience)
\begin{multline}
H_Q =  -\frac{\tilde{\varepsilon}_Z}{2} \sigma_z + \frac{1}{2}g\mu_B b x_0 \\ \times \left(  \eta \cos(\omega_d t) \sigma_x + \sin(\omega_d t) \sigma_y \right),
\end{multline}
where $\eta = \omega_d^2 /(\omega_x^2-\omega_d^2)$.
This Hamiltonian describes a pseudospin subject to a magnetic field rotating on an ellipse. In the rotating frame, we obtain
\begin{equation}
\tilde{H}_Q=\gamma \sigma_x,
\end{equation}
where $\gamma =\frac{1}{2}g\mu_Bbx_0 \frac{\eta+1}{2}$ by enforcing the resonance condition $\hbar\omega_d=\tilde{\varepsilon}_Z$ and ignoring the fast rotating term $\propto \begin{pmatrix} 0 & e^{-2i\omega_d t} \\ e^{2i\omega_d t} & 0\end{pmatrix}$. Finally, we obtain Eq.~\ref{eq:f_rabi} as $t_{\pi}=\pi\hbar/2\gamma$, with adiabatic and nonadiabatic contributions identified.

\section{Three electron EDSR in harmonic confinement}
\label{sec:appEDSR3e}
In this section, we extend the discussion to a 3-electron quantum dot.  Although, we study a (quasi) two dimensional case here we still only consider the effect of $B_0$ on spin, and neglect the inhomogeneous magnetic field in $x$, due to the strong confinement in $z$. 
The Hamiltonian reads as
\begin{equation}
\bar{H}=\bar{H}_{0}+\sum_{i=1}^{3}(H_{B_x}^{(i)}+H_{B_z}^{(i)}+H_t^{(i)}),
\end{equation}
where
\begin{multline}
\bar{H}_{0}=\sum_{i=1}^3 \left ( \frac{p_{x_i}^2}{2m^*} + \frac{1}{2}m^*\omega_x^2x_i^2 + \frac{p_{y_i}^2}{2m^*}  + \frac{1}{2}m^*\omega_y^2y_i^2 \right ) \\ + \sum_{i>j}^{3} \frac{e^2}{4 \pi \epsilon r_{ij}},
\end{multline}
with $r_{ij} \equiv |\bm{r}_i-\bm{r}_j|$.
We have the FCI solutions that obey
\begin{equation}
\bar{H}_{0} \ket{\Psi_n,s_x} = \bar{\varepsilon}_n \ket{\Psi_n,s_x},
\end{equation}
and we also assume
\begin{equation}
\left(\bar{H}_{0}+ \sum_{i=1}^{3}H_{B_x}^{(i)} \right ) \ket{\Psi_n,s_x} = \bar{\varepsilon}_n^{(s_x)} \ket{\Psi_n,s_x},
\end{equation}
where $\bar{\varepsilon}_n^{(s_x)}=\bar{\varepsilon_n}- \frac{1}{2}g\mu_B B_0 s_x $.

Assuming that the ground state of the unperturbed system has total spin $1/2$ (strong enough confinement), now we have the following qubit states:
\begin{subequations}
\label{eq:p_exp}
\begin{multline}
\ket{\bar{0}}\approx \ket{\Psi_0,+1}-\frac{1}{2}g \mu_B b \sum_{n=0}^{\infty} \frac{\bar{\Xi}_{n,0}^{-1,1} }{\bar{\varepsilon}_0^{(+1)}-\bar{\varepsilon}_n^{(-1)}} \ket{\Psi_n,-1} \\ +\sum_{m=0}^{\infty}c_m\ket{\Psi_m,+3},
\end{multline}

\begin{multline}
\ket{\bar{1}}\approx \ket{\Psi_0,-1}-\frac{1}{2}g \mu_B b \sum_{n=0}^{\infty} \frac{\bar{\Xi}_{n,0}^{1,-1} } {\bar{\varepsilon}_0^{(-1)}-\bar{\varepsilon}_n^{(+1)}} \ket{\Psi_n,+1} \\ +\sum_{m=0}^{\infty}c^\prime_m\ket{\Psi_m,-3},
\end{multline}
where
\begin{equation}
\bar{\Xi}_{n,n^\prime}^{s_x,s_x^\prime} \equiv \bra{\Psi_n,s_x} x_1 \sigma_{z_1} + x_2 \sigma_{z_2} + x_3 \sigma_{z_3} \ket{\Psi_{n^\prime},s_x^\prime}.
\end{equation}
\end{subequations}
Next, we need to calculate $\bar{V}_{01} \equiv e E_0 \bra{\bar{0}} x_1+x_2+x_3 \ket{\bar{1}}$. Note that $\ket{\Psi_m,\pm3}$ corrections above can be ignored at this point as they do not get coupled by the electric field. Moreover, we can observe that only the corrections that have a single excitation in $x$ the component of the center of mass motion get coupled by the electric field. To see this first we describe the system in center-of-mass and internal coordinates. A particular choice for the coordinate transformation is the following~\cite{Ruan:1995p7942}:
\begin{subequations}
\begin{equation}
\bm{R}=\frac{\bm{r}_1+\bm{r}_2+\bm{r}_3}{3},
\end{equation}

\begin{equation}
\bm{\xi}=\bm{r}_1-\bm{r}_2
\end{equation}

\begin{equation}
\bm{\eta}=\bm{r}_3-(\bm{r}_1+\bm{r}_2)/2.
\end{equation}
\end{subequations}
This transformation separates the Hamiltonian into
\begin{equation}
\bar{H}_0=H_{CM}+H_I,
\end{equation}
where
\begin{subequations}
\begin{equation}
H_{CM}=\frac{p_{R_x}^2}{2M}+\frac{p_{R_y}^2}{2M}+\frac{1}{2}M\omega_x^2 R_x^2+\frac{1}{2}M\omega_y^2 R_y^2,
\end{equation}
\begin{multline}
H_{I}=\frac{p_{\xi_x}^2}{2\mu_{\xi}}+\frac{p_{\xi_y}^2}{2\mu_{\xi}}+\frac{p_{\eta_x}^2}{2\mu_{\eta}}+\frac{p_{\eta_y}^2}{2\mu_{\eta}} \\ +\sum_{i>j}\left ( \frac{1}{6}m^*\omega_x^2x_{ij}^2+\frac{1}{6}m^*\omega_y^2y_{ij}^2+\frac{e^2}{4 \pi \epsilon r_{ij}} \right ),
\end{multline}
\end{subequations}
with $M=3m^*$, $\mu_{\xi}=m^*/2$, and $\mu_{\eta}=2m^*/3$.  We let
\begin{equation}
H_{I} \ket{\zeta_{n_I}} = \varepsilon_{n_I} \ket{\zeta_{n_I}}.
\end{equation}
The center of mass Hamiltonian is also separable in $x$ and $y$. Therefore, we can write
\begin{equation}
H_{CM} \ket{\bar{\phi}_{n_x,n_y}}=\varepsilon_{n_x,n_y} \ket{\bar{\phi}_{n_x,n_y} },
\end{equation}
where $\varepsilon_{n_x,n_y} =\hbar \omega_x (n_x +1/2)+\hbar \omega_y (n_y +1/2)$, $\bar{\phi}_{n_x,n_y} \equiv \phi_{n_x}(R_x,M) \phi_{n_y}(R_y,M)$,
and obtain

\begin{widetext}
\begin{equation}
\label{eq:x_couple}
\begin{aligned}
\bra{\Psi_{0,0,n_I},s_x} x_1+x_2+x_3 \ket{\Psi_{n_x,n_y,n_I^\prime},s_x} & =  \bra{\phi_{0}(R_x,M) \phi_{0}(R_y,M) \zeta_{n_I}(\bm{\xi},\bm{\eta})} 3R_x \ket{\phi_{n_x}(R_x,M) \phi_{n_y}(R_y,M) \zeta_{n_I^\prime}(\bm{\xi},\bm{\eta})} \\
&=3\bra{\phi_{0}(R_x,3m^*)} R_x \ket{\phi_{n_x}(R_x,3m^*)} \delta_{0,n_y} \delta_{n_I,n_{I^\prime}} \\
&=3 \sqrt{n_x} \delta_{0,n_x -1} \frac{a_0}{\sqrt{3}\sqrt{2}} \delta_{0,n_y} \delta_{n_I,n_{I^\prime}}.
\end{aligned}
\end{equation}

\begin{table*}[h]
\caption{Spin states of three electrons is composed of a $S=3\hbar/2$ quadruplet ($Q$) and two $S=\hbar/2$ doublets (${D_a}$ and ${D_b}$).}
\label{tab:spinTable}
\begin{tabular}{@{}llllllllllllc@{}}
Label             &  &  &  & $S$                  &  &  &  & $S_x$ &  &  &  & Spin state                                                                                                                         \\
                  &  &  &  &                      &  &  &  &       &  &  &  & \multicolumn{1}{l}{}                                                                                                               \\
${D_a^-}$  &  &  &  & \multirow{2}{*}{$\hbar/2$} &  &  &  & $-\hbar/2$   &  &  &  & $(\ket{- + -} -  \ket{+ - -})/\sqrt{2} $                                             \\
${D_a^+}$  &  &  &  &                      &  &  &  & $+\hbar/2$  &  &  &  & $( \ket{+ - +} - \ket{- + +})/\sqrt{2} $                                         \\
                  &  &  &  &                      &  &  &  &       &  &  &  & \multicolumn{1}{l}{}                                                                                                               \\
${D_b^-}$  &  &  &  & \multirow{2}{*}{$\hbar$/2} &  &  &  & $-\hbar$/2   &  &  &  & $(2 \ket{- - +} - \ket{- + -} -  \ket{+ - -})/\sqrt{6}$       \\
${D_b^+}$  &  &  &  &                      &  &  &  & $+\hbar$/2  &  &  &  &  $(2 \ket{+ + -} - \ket{+ - +} -  \ket{- + +})/\sqrt{6}$\\
\\
${Q^{--}}$ &  &  &  & \multirow{4}{*}{3$\hbar$/2} &  &  &  & $-3\hbar/2$   &  &  &  & $\ket{- - -}$                                                                                                 \\
${Q^{-}}$  &  &  &  &                      &  &  &  & $-\hbar/2$   &  &  &  & $(\ket{- - +} + \ket{- + -} + \ket{+ - -})/\sqrt{3}$        \\
${Q^{+}}$  &  &  &  &                      &  &  &  & $+\hbar/2$  &  &  &  &  $( \ket{+ + -} +  \ket{+ - +} + \ket{- + +})/\sqrt{3}$   \\
${Q^{++}}$ &  &  &  &                      &  &  &  & $+3\hbar/2$  &  &  &  & $\ket{+ + +}$                                                                                           \\

\end{tabular}
\end{table*}

\end{widetext}
So, similar to the single electron case, the main coupling between the qubit states due to the electric field comes from the corrections that lie $\hbar \omega_x$ above the ground state, \textit{i.e.} $n_x=1$. However, the coupling is $\sqrt{3}$ times as strong as in the single-electron case. 

Now, to look at how the weight of these corrections change going from single- to three-electron case, we consider
\begin{equation}
\label{eq:bCouple}
\bra{\Psi_{0,0,n_I},\pm1} \sigma_{z_1}x_1+\sigma_{z_2}x_2+\sigma_{z_3}x_3 \ket{\Psi_{1,0,n_I},\mp1}.
\end{equation}

\begin{widetext}

Since these states are $S=1/2$ doublets (Total spin eigenstates are shown in Table \ref{tab:spinTable}), we write them as follows:
\begin{equation}
\begin{aligned}
\ket{\Psi_{0,0,n_I},\pm1} &=\alpha \ket{\bar{\phi}_{0,0}} \ket{\zeta_a} \ket{D_a^{\pm}}+\beta \ket{\bar{\phi}_{0,0}} \ket{\zeta_b} \ket{D_b^{\pm}}
\\
\ket{\Psi_{1,0,n_I},\pm1} &=\alpha \ket{\bar{\phi}_{1,0}} \ket{\zeta_a} \ket{D_a^{\pm}}+\beta \ket{\bar{\phi}_{1,0}} \ket{\zeta_b} \ket{D_b^{\pm}}.
\end{aligned}
\end{equation}

To evaluate the expression in (\ref{eq:bCouple}), we also need the matrix elements of $\sigma_{z_i}$ in $\{D_a^{-}, D_a^{+},D_b^{-},D_b^{+} \}$ basis:

\begin{subequations}
\begin{minipage}{.33333\textwidth}
  \begin{equation}
    \sigma_{z_1}=
      \begin{pmatrix}
0 & 0 & 0 & \sfrac{1}{\sqrt{3}} \\
0 & 0 &  \sfrac{1}{\sqrt{3}} & 0 \\
0 &  \sfrac{1}{\sqrt{3}} & 0 &\sfrac{-2}{3} \\
\sfrac{1}{\sqrt{3}} & 0  & \sfrac{-2}{3} & 0
\end{pmatrix}  
\end{equation}
\end{minipage}
\begin{minipage}{.33333\textwidth}
  \begin{equation}
    \sigma_{z_2}=
      \begin{pmatrix}
0 & 0 & 0 & \sfrac{-1}{\sqrt{3}} \\
0 & 0 &  \sfrac{-1}{\sqrt{3}} & 0 \\
0 &  \sfrac{-1}{\sqrt{3}} & 0 &\sfrac{-2}{3} \\
\sfrac{-1}{\sqrt{3}} & 0  & \sfrac{-2}{3} & 0
\end{pmatrix}  
\end{equation}
\end{minipage}
\begin{minipage}{.33333\textwidth}
  \begin{equation}
    \sigma_{z_3}=
      \begin{pmatrix}
0 & -1 & 0 & 0 \\
-1 & 0 &  0 & 0 \\
0 &  0 & 0 &\sfrac{1}{3} \\
0 & 0  & \sfrac{1}{3} & 0
\end{pmatrix}  .
\end{equation}
\end{minipage}
\end{subequations}
With these, we obtain
\begin{multline}
\label{eq:x_sigma_couple}
\bra{\Psi_{0,0,n_I},\pm1} \sigma_{z_1}x_1+\sigma_{z_2}x_2+\sigma_{z_3}x_3 \ket{\Psi_{1,0,n_I},\mp1}\\=-\alpha^2 \bra{\bar{\phi}_{0,0}\zeta_a} x_3 \ket{\bar{\phi}_{1,0}\zeta_a} +\beta^2 \bra{\bar{\phi}_{0,0}\zeta_b} -\frac{2}{3}x_1 -\frac{2}{3}x_2 + \frac{1}{3}x_3 \ket{\bar{\phi}_{1,0}\zeta_b} \\
+\alpha \beta \left [ \frac{1}{\sqrt{3}} \bra{\bar{\phi}_{0,0}\zeta_a} x_1 \ket{\bar{\phi}_{1,0}\zeta_b}-\frac{1}{\sqrt{3}} \bra{\bar{\phi}_{0,0}\zeta_a} x_2 \ket{\bar{\phi}_{1,0}\zeta_b} + \frac{1}{\sqrt{3}} \bra{\bar{\phi}_{0,0}\zeta_b} x_1 \ket{\bar{\phi}_{1,0}\zeta_a}-\frac{1}{\sqrt{3}} \bra{\bar{\phi}_{0,0}\zeta_b} x_2 \ket{\bar{\phi}_{1,0}\zeta_a} \right ] \\
=\alpha^2 \bra{\bar{\phi}_{0,0}\zeta_a} -R_x-2\eta_x/3 \ket{\bar{\phi}_{1,0}\zeta_a}+\beta^2  \bra{\bar{\phi}_{0,0}\zeta_a} -R_x+2\eta_x/3 \ket{\bar{\phi}_{1,0}\zeta_a} \\ 
+\frac{\alpha\beta}{\sqrt{3}} \bra{\bar{\phi}_{0,0}\zeta_a} \xi_x \ket{\bar{\phi}_{1,0}\zeta_b}+\frac{\alpha\beta}{\sqrt{3}}  \bra{\bar{\phi}_{0,0}\zeta_b} \xi_x \ket{\bar{\phi}_{1,0}\zeta_a}  \\
=-\bra{\bar{\phi}_{0,0}} R_x \ket{\bar{\phi}_{1,0}} 
=-\frac{a_0}{\sqrt{3} \sqrt{2}}
\end{multline}

Using Eq.~\ref{eq:x_couple} and Eq.~\ref{eq:x_sigma_couple} we obtain
\begin{multline}
\bar{V}_{01} =\frac{1}{2}eE_0g\mu_Bb \bra{\Psi_{0,0,n_I},\pm1} \sigma_{z_1}x_1+\sigma_{z_2}x_2+\sigma_{z_3}x_3 \ket{\Psi_{1,0,n_I},\mp1} \bra{\Psi_{0,0,n_I},\pm1} x_1+x_2+x_3 \ket{\Psi_{1,0,n_I},\pm1}  \left ( \frac{2 \hbar \omega_x}{(\hbar \omega_x)^2-\varepsilon_Z^2} \right ) \\
=\frac{1}{2} e E_0 g \mu_B b \left ( -\frac{a_0}{\sqrt{6}} \right ) \left ( 3\frac{a_0}{\sqrt{6}} \right ) \left ( \frac{2 \hbar \omega_x}{(\hbar \omega_x)^2-\varepsilon_Z^2} \right ) \\ = -\frac{1}{2}eE_0g \mu_B b a_0^2 \left ( \frac{ \hbar \omega}{(\hbar \omega)^2-\varepsilon_Z^2} \right )
\end{multline}
Hence, $|\bar{V}_{01}|=|V_{01}|$ and there is no difference in $t_{\pi}$, to the first order, between single- and three-electron cases.
\end{widetext}

\section{Double dots}
\label{Sec:DoubleDot}
It is known that single-electron EDSR in a double dot, i.e. ``flopping mode" qubit, yields a very large response with Rabi frequencies that are orders of magnitude larger than single dot implementations~\cite{Croot:2020p012006}.  This remains true in the case of three electrons, as we demonstrate in this subsection. 

\begin{figure}[h]
\includegraphics[width=3.2 in]{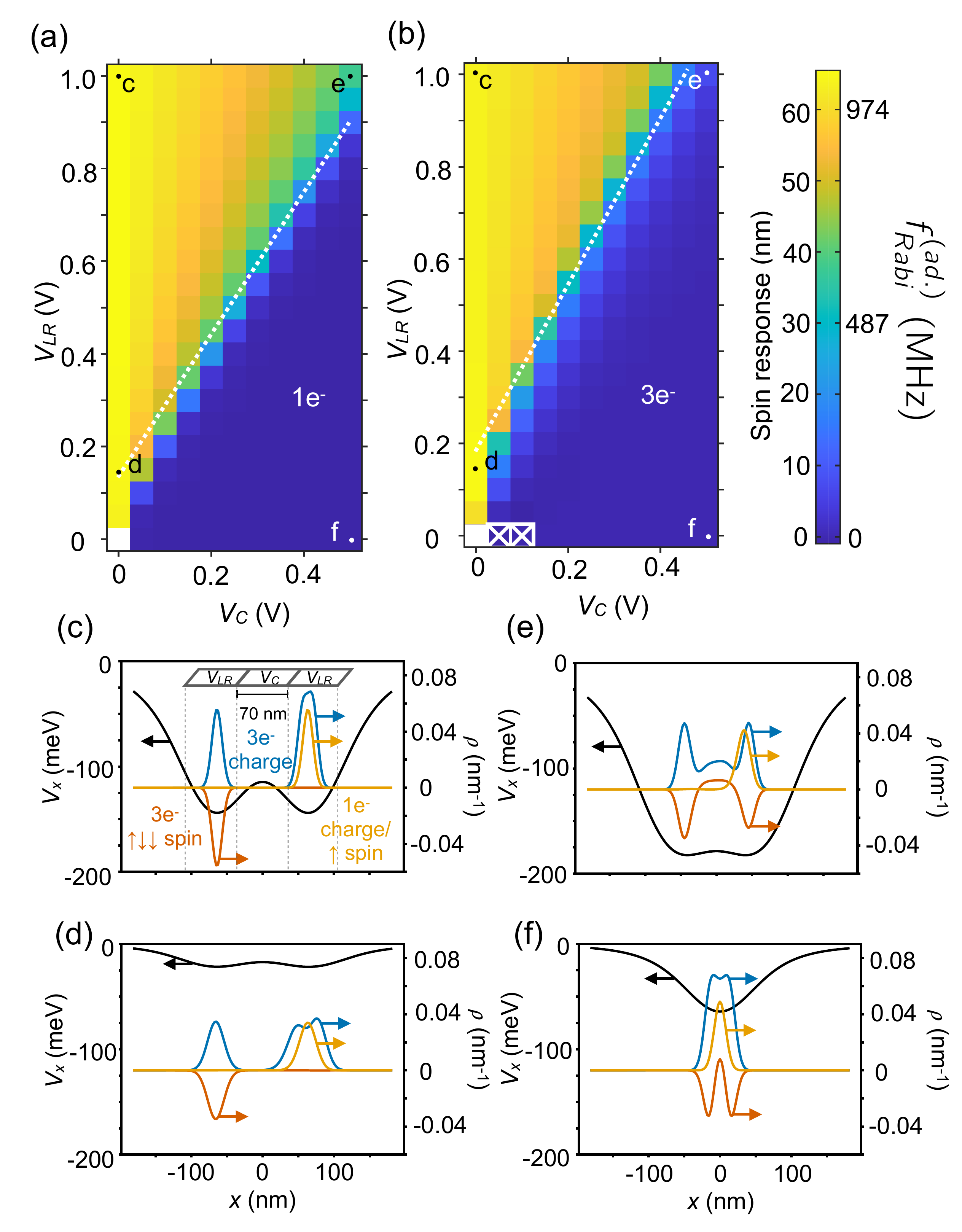}
\caption{Spin response of electron(s) in a confinement potential induced by three square gates(geometry shown in the inset to (c)) to the electric field $E_0$ (a) Single-electron charge/spin center as functions of the center and edge gate potentials $V_C$ and $V_{LR}$. Geometry of the gates is shown schematically in the inset of (c). (b) Three-electron spin center as functions of $V_C$ and $V_{LR}$. White crosses indicate $S=3/2$ ground states. 
Adiabatic estimates of $f_{Rabi}$ corresponding to the spin response are also shown on the color scale. (c-f) Confinement potentials (black), single-electron charge/spin (orange), three-electron charge (blue) and spin (red) densities are shown for four selected voltage pairs $(V_C,V_{LR})$ marked in (a), (b).
}
\label{fig:doubleDot}
\end{figure}

We consider lateral potentials generated by three 70 nm $\times$ 70 nm square gates~\cite{Anderson:2022p065123} shown schematically in the inset of Fig.~\ref{fig:doubleDot}(c) that are 70~nm above a two-dimensional electron gas. 
%The two-dimensional electron gas is assumed to be formed 70 nm below the gates. 
The spin responses to electric field $\bar{E}_0$ as a function of the edge and center biases, $V_{LR}$ and $V_C$, for one and three electrons are reported in Fig.~\ref{fig:doubleDot}(a) and (b), respectively. 
Potential profiles in $x$ together with charge and spin densities corresponding to four $(V_{LR},V_C)$ pairs marked with letters $c$, $d$, $e$, $f$ are also given in Fig.~\ref{fig:doubleDot}(c-f). 
It is seen that double-dot responses, such as points $c$ and $d$, are similarly much larger than single-dot responses, such as points $f$, for both single- and three-electron cases. A single electron in a double dot EDSR simply hops between the two dots as driven by the AC field, as indicated by its density (yellow line) in Fig.~\ref{fig:doubleDot}(c), (d) (density plots (c-f) corresponding to $-\bar{E}_0$ are just the mirror symmetries around $x=0$). The three-electron situation is similar here, except that the spin oscillations are out of phase with the single-electron case,  as two of the electrons constantly occupy each dot while the third one hops between them: A spin singlet forms in the doubly-occupied dot as indicated by the larger charge density (blue line) accompanied by a vanishing spin density (red line) in the right dot. The point $f$ corresponds to a soft anharmonic confinement that is discussed in Sec.~\ref{Sec:anharmonic} as the confinement profile in Fig.~\ref{fig:doubleDot}(f) indicates. Due to the constant occupation of both dots, the transition from the low to high response regimes (from point $f$ to $c$) requires higher barriers (lower $V_C$) / deeper dots (higher $V_{LR}$) for three electrons. This can be seen in Fig.~\ref{fig:doubleDot}(e): while the single-electron density is localized in the right dot, the three-electron spin density is distributed between both dots. As a result, the point $e$ yields high response for a single-electron but low response for three electrons. The white dotted lines in Fig.~\ref{fig:doubleDot}(a-b) are drawn as guides to the eye to indicate this transition: in Fig.~\ref{fig:doubleDot}(b) the line is located higher than in Fig.~\ref{fig:doubleDot}(a).

In summary, adding more electrons to the double dot changes the voltages at which the response is minimized, but it does not lead to an enhancement of the Rabi frequency.

\bibliography{text}

\begin{thebibliography}{31}%
\makeatletter
\providecommand \@ifxundefined [1]{%
 \@ifx{#1\undefined}
}%
\providecommand \@ifnum [1]{%
 \ifnum #1\expandafter \@firstoftwo
 \else \expandafter \@secondoftwo
 \fi
}%
\providecommand \@ifx [1]{%
 \ifx #1\expandafter \@firstoftwo
 \else \expandafter \@secondoftwo
 \fi
}%
\providecommand \natexlab [1]{#1}%
\providecommand \enquote  [1]{``#1''}%
\providecommand \bibnamefont  [1]{#1}%
\providecommand \bibfnamefont [1]{#1}%
\providecommand \citenamefont [1]{#1}%
\providecommand \href@noop [0]{\@secondoftwo}%
\providecommand \href [0]{\begingroup \@sanitize@url \@href}%
\providecommand \@href[1]{\@@startlink{#1}\@@href}%
\providecommand \@@href[1]{\endgroup#1\@@endlink}%
\providecommand \@sanitize@url [0]{\catcode `\\12\catcode `\$12\catcode
  `\&12\catcode `\#12\catcode `\^12\catcode `\_12\catcode `\%12\relax}%
\providecommand \@@startlink[1]{}%
\providecommand \@@endlink[0]{}%
\providecommand \url  [0]{\begingroup\@sanitize@url \@url }%
\providecommand \@url [1]{\endgroup\@href {#1}{\urlprefix }}%
\providecommand \urlprefix  [0]{URL }%
\providecommand \Eprint [0]{\href }%
\providecommand \doibase [0]{http://dx.doi.org/}%
\providecommand \selectlanguage [0]{\@gobble}%
\providecommand \bibinfo  [0]{\@secondoftwo}%
\providecommand \bibfield  [0]{\@secondoftwo}%
\providecommand \translation [1]{[#1]}%
\providecommand \BibitemOpen [0]{}%
\providecommand \bibitemStop [0]{}%
\providecommand \bibitemNoStop [0]{.\EOS\space}%
\providecommand \EOS [0]{\spacefactor3000\relax}%
\providecommand \BibitemShut  [1]{\csname bibitem#1\endcsname}%
\let\auto@bib@innerbib\@empty
%</preamble>
\bibitem [{\citenamefont {Koppens}\ \emph {et~al.}(2006)\citenamefont
  {Koppens}, \citenamefont {Buizert}, \citenamefont {Tielrooij}, \citenamefont
  {Vink}, \citenamefont {Nowack}, \citenamefont {Meunier}, \citenamefont
  {Kouwenhoven},\ and\ \citenamefont {Vandersypen}}]{Koppens:2006p766}%
  \BibitemOpen
  \bibfield  {author} {\bibinfo {author} {\bibfnamefont {F.~H.~L.}\
  \bibnamefont {Koppens}}, \bibinfo {author} {\bibfnamefont {C.}~\bibnamefont
  {Buizert}}, \bibinfo {author} {\bibfnamefont {K.~J.}\ \bibnamefont
  {Tielrooij}}, \bibinfo {author} {\bibfnamefont {I.~T.}\ \bibnamefont {Vink}},
  \bibinfo {author} {\bibfnamefont {K.~C.}\ \bibnamefont {Nowack}}, \bibinfo
  {author} {\bibfnamefont {T.}~\bibnamefont {Meunier}}, \bibinfo {author}
  {\bibfnamefont {L.~P.}\ \bibnamefont {Kouwenhoven}}, \ and\ \bibinfo {author}
  {\bibfnamefont {L.~M.~K.}\ \bibnamefont {Vandersypen}},\ }\href {\doibase
  10.1038/nature05065} {\bibfield  {journal} {\bibinfo  {journal} {Nature}\
  }\textbf {\bibinfo {volume} {442}},\ \bibinfo {pages} {766} (\bibinfo {year}
  {2006})}\BibitemShut {NoStop}%
\bibitem [{\citenamefont {Veldhorst}\ \emph {et~al.}(2014)\citenamefont
  {Veldhorst}, \citenamefont {Hwang}, \citenamefont {Yang}, \citenamefont
  {Leenstra}, \citenamefont {de~Ronde}, \citenamefont {Dehollain},
  \citenamefont {Muhonen}, \citenamefont {Hudson}, \citenamefont {Itoh},
  \citenamefont {Morello},\ and\ \citenamefont {Dzurak}}]{Veldhorst:2014p981}%
  \BibitemOpen
  \bibfield  {author} {\bibinfo {author} {\bibfnamefont {M.}~\bibnamefont
  {Veldhorst}}, \bibinfo {author} {\bibfnamefont {J.~C.~C.}\ \bibnamefont
  {Hwang}}, \bibinfo {author} {\bibfnamefont {C.~H.}\ \bibnamefont {Yang}},
  \bibinfo {author} {\bibfnamefont {A.~W.}\ \bibnamefont {Leenstra}}, \bibinfo
  {author} {\bibfnamefont {B.}~\bibnamefont {de~Ronde}}, \bibinfo {author}
  {\bibfnamefont {J.~P.}\ \bibnamefont {Dehollain}}, \bibinfo {author}
  {\bibfnamefont {J.~T.}\ \bibnamefont {Muhonen}}, \bibinfo {author}
  {\bibfnamefont {F.~E.}\ \bibnamefont {Hudson}}, \bibinfo {author}
  {\bibfnamefont {K.~M.}\ \bibnamefont {Itoh}}, \bibinfo {author}
  {\bibfnamefont {A.}~\bibnamefont {Morello}}, \ and\ \bibinfo {author}
  {\bibfnamefont {A.~S.}\ \bibnamefont {Dzurak}},\ }\href {\doibase
  10.1038/nnano.2014.216} {\bibfield  {journal} {\bibinfo  {journal} {Nature
  Nanotechnol.}\ }\textbf {\bibinfo {volume} {9}},\ \bibinfo {pages} {981}
  (\bibinfo {year} {2014})}\BibitemShut {NoStop}%
\bibitem [{\citenamefont {Pioro-Ladri\`{e}re}\ \emph
  {et~al.}(2008)\citenamefont {Pioro-Ladri\`{e}re}, \citenamefont {Obata},
  \citenamefont {Tokura}, \citenamefont {Shin}, \citenamefont {Kubo},
  \citenamefont {Yoshida}, \citenamefont {Taniyama},\ and\ \citenamefont
  {Tarucha}}]{PioroLadriere:2008p776}%
  \BibitemOpen
  \bibfield  {author} {\bibinfo {author} {\bibfnamefont {M.}~\bibnamefont
  {Pioro-Ladri\`{e}re}}, \bibinfo {author} {\bibfnamefont {T.}~\bibnamefont
  {Obata}}, \bibinfo {author} {\bibfnamefont {Y.}~\bibnamefont {Tokura}},
  \bibinfo {author} {\bibfnamefont {Y.-S.}\ \bibnamefont {Shin}}, \bibinfo
  {author} {\bibfnamefont {T.}~\bibnamefont {Kubo}}, \bibinfo {author}
  {\bibfnamefont {K.}~\bibnamefont {Yoshida}}, \bibinfo {author} {\bibfnamefont
  {T.}~\bibnamefont {Taniyama}}, \ and\ \bibinfo {author} {\bibfnamefont
  {S.}~\bibnamefont {Tarucha}},\ }\href {\doibase doi:10.1038/nphys1053}
  {\bibfield  {journal} {\bibinfo  {journal} {Nat. Phys.}\ }\textbf {\bibinfo
  {volume} {4}},\ \bibinfo {pages} {776} (\bibinfo {year} {2008})}\BibitemShut
  {NoStop}%
\bibitem [{\citenamefont {Obata}\ \emph {et~al.}(2010)\citenamefont {Obata},
  \citenamefont {Pioro-Ladriere}, \citenamefont {Tokura}, \citenamefont {Shin},
  \citenamefont {Kubo}, \citenamefont {Yoshida}, \citenamefont {Taniyama},\
  and\ \citenamefont {Tarucha}}]{Obata:2010p2612}%
  \BibitemOpen
  \bibfield  {author} {\bibinfo {author} {\bibfnamefont {T.}~\bibnamefont
  {Obata}}, \bibinfo {author} {\bibfnamefont {M.}~\bibnamefont
  {Pioro-Ladriere}}, \bibinfo {author} {\bibfnamefont {Y.}~\bibnamefont
  {Tokura}}, \bibinfo {author} {\bibfnamefont {Y.-S.}\ \bibnamefont {Shin}},
  \bibinfo {author} {\bibfnamefont {T.}~\bibnamefont {Kubo}}, \bibinfo {author}
  {\bibfnamefont {K.}~\bibnamefont {Yoshida}}, \bibinfo {author} {\bibfnamefont
  {T.}~\bibnamefont {Taniyama}}, \ and\ \bibinfo {author} {\bibfnamefont
  {S.}~\bibnamefont {Tarucha}},\ }\href {\doibase 10.1103/PhysRevB.81.085317}
  {\bibfield  {journal} {\bibinfo  {journal} {Phys. Rev. B}\ }\textbf {\bibinfo
  {volume} {81}},\ \bibinfo {pages} {085317} (\bibinfo {year}
  {2010})}\BibitemShut {NoStop}%
\bibitem [{\citenamefont {Kawakami}\ \emph {et~al.}(2014)\citenamefont
  {Kawakami}, \citenamefont {Scarlino}, \citenamefont {Ward}, \citenamefont
  {Braakman}, \citenamefont {Savage}, \citenamefont {Lagally}, \citenamefont
  {Friesen}, \citenamefont {Coppersmith}, \citenamefont {Eriksson},\ and\
  \citenamefont {Vandersypen}}]{Kawakami:2014p666}%
  \BibitemOpen
  \bibfield  {author} {\bibinfo {author} {\bibfnamefont {E.}~\bibnamefont
  {Kawakami}}, \bibinfo {author} {\bibfnamefont {P.}~\bibnamefont {Scarlino}},
  \bibinfo {author} {\bibfnamefont {D.~R.}\ \bibnamefont {Ward}}, \bibinfo
  {author} {\bibfnamefont {F.~R.}\ \bibnamefont {Braakman}}, \bibinfo {author}
  {\bibfnamefont {D.~E.}\ \bibnamefont {Savage}}, \bibinfo {author}
  {\bibfnamefont {M.~G.}\ \bibnamefont {Lagally}}, \bibinfo {author}
  {\bibfnamefont {M.}~\bibnamefont {Friesen}}, \bibinfo {author} {\bibfnamefont
  {S.~N.}\ \bibnamefont {Coppersmith}}, \bibinfo {author} {\bibfnamefont
  {M.~A.}\ \bibnamefont {Eriksson}}, \ and\ \bibinfo {author} {\bibfnamefont
  {L.~M.~K.}\ \bibnamefont {Vandersypen}},\ }\href {\doibase
  10.1038/nnano.2014.153} {\bibfield  {journal} {\bibinfo  {journal} {Nat.
  Nanotechnol.}\ }\textbf {\bibinfo {volume} {9}},\ \bibinfo {pages} {666}
  (\bibinfo {year} {2014})}\BibitemShut {NoStop}%
\bibitem [{\citenamefont {Takeda}\ \emph {et~al.}(2016)\citenamefont {Takeda},
  \citenamefont {Kamioka}, \citenamefont {Otsuka}, \citenamefont {Yoneda},
  \citenamefont {Nakajima}, \citenamefont {Delbecq}, \citenamefont {Amaha},
  \citenamefont {Allison}, \citenamefont {Kodera}, \citenamefont {Oda},\ and\
  \citenamefont {Tarucha}}]{Takeda:2016p1600694}%
  \BibitemOpen
  \bibfield  {author} {\bibinfo {author} {\bibfnamefont {K.}~\bibnamefont
  {Takeda}}, \bibinfo {author} {\bibfnamefont {J.}~\bibnamefont {Kamioka}},
  \bibinfo {author} {\bibfnamefont {T.}~\bibnamefont {Otsuka}}, \bibinfo
  {author} {\bibfnamefont {J.}~\bibnamefont {Yoneda}}, \bibinfo {author}
  {\bibfnamefont {T.}~\bibnamefont {Nakajima}}, \bibinfo {author}
  {\bibfnamefont {M.~R.}\ \bibnamefont {Delbecq}}, \bibinfo {author}
  {\bibfnamefont {S.}~\bibnamefont {Amaha}}, \bibinfo {author} {\bibfnamefont
  {G.}~\bibnamefont {Allison}}, \bibinfo {author} {\bibfnamefont
  {T.}~\bibnamefont {Kodera}}, \bibinfo {author} {\bibfnamefont
  {S.}~\bibnamefont {Oda}}, \ and\ \bibinfo {author} {\bibfnamefont
  {S.}~\bibnamefont {Tarucha}},\ }\href {\doibase 10.1126/sciadv.1600694}
  {\bibfield  {journal} {\bibinfo  {journal} {Sci. Adv.}\ }\textbf {\bibinfo
  {volume} {2}},\ \bibinfo {pages} {1600694} (\bibinfo {year}
  {2016})}\BibitemShut {NoStop}%
\bibitem [{\citenamefont {Watson}\ \emph {et~al.}(2018)\citenamefont {Watson},
  \citenamefont {Philips}, \citenamefont {Kawakami}, \citenamefont {Ward},
  \citenamefont {Scarlino}, \citenamefont {Veldhorst}, \citenamefont {Savage},
  \citenamefont {Lagally}, \citenamefont {Friesen}, \citenamefont
  {Coppersmith}, \citenamefont {Eriksson},\ and\ \citenamefont
  {Vandersypen}}]{Watson:2018p633}%
  \BibitemOpen
  \bibfield  {author} {\bibinfo {author} {\bibfnamefont {T.~F.}\ \bibnamefont
  {Watson}}, \bibinfo {author} {\bibfnamefont {S.~G.~J.}\ \bibnamefont
  {Philips}}, \bibinfo {author} {\bibfnamefont {E.}~\bibnamefont {Kawakami}},
  \bibinfo {author} {\bibfnamefont {D.~R.}\ \bibnamefont {Ward}}, \bibinfo
  {author} {\bibfnamefont {P.}~\bibnamefont {Scarlino}}, \bibinfo {author}
  {\bibfnamefont {M.}~\bibnamefont {Veldhorst}}, \bibinfo {author}
  {\bibfnamefont {D.~E.}\ \bibnamefont {Savage}}, \bibinfo {author}
  {\bibfnamefont {M.~G.}\ \bibnamefont {Lagally}}, \bibinfo {author}
  {\bibfnamefont {M.}~\bibnamefont {Friesen}}, \bibinfo {author} {\bibfnamefont
  {S.~N.}\ \bibnamefont {Coppersmith}}, \bibinfo {author} {\bibfnamefont
  {M.~A.}\ \bibnamefont {Eriksson}}, \ and\ \bibinfo {author} {\bibfnamefont
  {L.~M.~K.}\ \bibnamefont {Vandersypen}},\ }\href {\doibase
  10.1038/nature25766} {\bibfield  {journal} {\bibinfo  {journal} {Nature}\
  }\textbf {\bibinfo {volume} {555}},\ \bibinfo {pages} {633} (\bibinfo {year}
  {2018})}\BibitemShut {NoStop}%
\bibitem [{\citenamefont {Yang}\ \emph {et~al.}(2020)\citenamefont {Yang},
  \citenamefont {Leon}, \citenamefont {Hwang}, \citenamefont {Saraiva},
  \citenamefont {Tanttu}, \citenamefont {Huang}, \citenamefont
  {Camirand~Lemyre}, \citenamefont {Chan}, \citenamefont {Tan}, \citenamefont
  {Hudson}, \citenamefont {Itoh}, \citenamefont {Morello}, \citenamefont
  {Pioro-Ladri{\`e}re}, \citenamefont {Laucht},\ and\ \citenamefont
  {Dzurak}}]{Yang:2020p350}%
  \BibitemOpen
  \bibfield  {author} {\bibinfo {author} {\bibfnamefont {C.~H.}\ \bibnamefont
  {Yang}}, \bibinfo {author} {\bibfnamefont {R.~C.~C.}\ \bibnamefont {Leon}},
  \bibinfo {author} {\bibfnamefont {J.~C.~C.}\ \bibnamefont {Hwang}}, \bibinfo
  {author} {\bibfnamefont {A.}~\bibnamefont {Saraiva}}, \bibinfo {author}
  {\bibfnamefont {T.}~\bibnamefont {Tanttu}}, \bibinfo {author} {\bibfnamefont
  {W.}~\bibnamefont {Huang}}, \bibinfo {author} {\bibfnamefont
  {J.}~\bibnamefont {Camirand~Lemyre}}, \bibinfo {author} {\bibfnamefont
  {K.~W.}\ \bibnamefont {Chan}}, \bibinfo {author} {\bibfnamefont {K.~Y.}\
  \bibnamefont {Tan}}, \bibinfo {author} {\bibfnamefont {F.~E.}\ \bibnamefont
  {Hudson}}, \bibinfo {author} {\bibfnamefont {K.~M.}\ \bibnamefont {Itoh}},
  \bibinfo {author} {\bibfnamefont {A.}~\bibnamefont {Morello}}, \bibinfo
  {author} {\bibfnamefont {M.}~\bibnamefont {Pioro-Ladri{\`e}re}}, \bibinfo
  {author} {\bibfnamefont {A.}~\bibnamefont {Laucht}}, \ and\ \bibinfo {author}
  {\bibfnamefont {A.~S.}\ \bibnamefont {Dzurak}},\ }\href {\doibase
  10.1038/s41586-020-2171-6} {\bibfield  {journal} {\bibinfo  {journal}
  {Nature}\ }\textbf {\bibinfo {volume} {580}},\ \bibinfo {pages} {350}
  (\bibinfo {year} {2020})}\BibitemShut {NoStop}%
\bibitem [{\citenamefont {Noiri}\ \emph {et~al.}(2022)\citenamefont {Noiri},
  \citenamefont {Takeda}, \citenamefont {Nakajima}, \citenamefont {Kobayashi},
  \citenamefont {Sammak}, \citenamefont {Scappucci},\ and\ \citenamefont
  {Tarucha}}]{Noiri:2022p338}%
  \BibitemOpen
  \bibfield  {author} {\bibinfo {author} {\bibfnamefont {A.}~\bibnamefont
  {Noiri}}, \bibinfo {author} {\bibfnamefont {K.}~\bibnamefont {Takeda}},
  \bibinfo {author} {\bibfnamefont {T.}~\bibnamefont {Nakajima}}, \bibinfo
  {author} {\bibfnamefont {T.}~\bibnamefont {Kobayashi}}, \bibinfo {author}
  {\bibfnamefont {A.}~\bibnamefont {Sammak}}, \bibinfo {author} {\bibfnamefont
  {G.}~\bibnamefont {Scappucci}}, \ and\ \bibinfo {author} {\bibfnamefont
  {S.}~\bibnamefont {Tarucha}},\ }\href {\doibase 10.1038/s41586-021-04182-y}
  {\bibfield  {journal} {\bibinfo  {journal} {Nature}\ }\textbf {\bibinfo
  {volume} {601}},\ \bibinfo {pages} {338} (\bibinfo {year}
  {2022})}\BibitemShut {NoStop}%
\bibitem [{\citenamefont {Leon}\ \emph {et~al.}(2020)\citenamefont {Leon},
  \citenamefont {Yang}, \citenamefont {Hwang}, \citenamefont {Lemyre},
  \citenamefont {Tanttu}, \citenamefont {Huang}, \citenamefont {Chan},
  \citenamefont {Tan}, \citenamefont {Hudson}, \citenamefont {Itoh},
  \citenamefont {Morello}, \citenamefont {Laucht}, \citenamefont
  {Pioro-Ladri{\`e}re}, \citenamefont {Saraiva},\ and\ \citenamefont
  {Dzurak}}]{Leon:2020p797}%
  \BibitemOpen
  \bibfield  {author} {\bibinfo {author} {\bibfnamefont {R.~C.~C.}\
  \bibnamefont {Leon}}, \bibinfo {author} {\bibfnamefont {C.~H.}\ \bibnamefont
  {Yang}}, \bibinfo {author} {\bibfnamefont {J.~C.~C.}\ \bibnamefont {Hwang}},
  \bibinfo {author} {\bibfnamefont {J.~C.}\ \bibnamefont {Lemyre}}, \bibinfo
  {author} {\bibfnamefont {T.}~\bibnamefont {Tanttu}}, \bibinfo {author}
  {\bibfnamefont {W.}~\bibnamefont {Huang}}, \bibinfo {author} {\bibfnamefont
  {K.~W.}\ \bibnamefont {Chan}}, \bibinfo {author} {\bibfnamefont {K.~Y.}\
  \bibnamefont {Tan}}, \bibinfo {author} {\bibfnamefont {F.~E.}\ \bibnamefont
  {Hudson}}, \bibinfo {author} {\bibfnamefont {K.~M.}\ \bibnamefont {Itoh}},
  \bibinfo {author} {\bibfnamefont {A.}~\bibnamefont {Morello}}, \bibinfo
  {author} {\bibfnamefont {A.}~\bibnamefont {Laucht}}, \bibinfo {author}
  {\bibfnamefont {M.}~\bibnamefont {Pioro-Ladri{\`e}re}}, \bibinfo {author}
  {\bibfnamefont {A.}~\bibnamefont {Saraiva}}, \ and\ \bibinfo {author}
  {\bibfnamefont {A.~S.}\ \bibnamefont {Dzurak}},\ }\href {\doibase
  10.1038/s41467-019-14053-w} {\bibfield  {journal} {\bibinfo  {journal}
  {Nature Communications}\ }\textbf {\bibinfo {volume} {11}},\ \bibinfo {pages}
  {797} (\bibinfo {year} {2020})}\BibitemShut {NoStop}%
\bibitem [{\citenamefont {Leon}\ \emph {et~al.}(2021)\citenamefont {Leon},
  \citenamefont {Yang}, \citenamefont {Hwang}, \citenamefont {Camirand~Lemyre},
  \citenamefont {Tanttu}, \citenamefont {Huang}, \citenamefont {Huang},
  \citenamefont {Hudson}, \citenamefont {Itoh}, \citenamefont {Laucht},
  \citenamefont {Pioro-Ladri{\`e}re}, \citenamefont {Saraiva},\ and\
  \citenamefont {Dzurak}}]{Leon:2021p3228}%
  \BibitemOpen
  \bibfield  {author} {\bibinfo {author} {\bibfnamefont {R.~C.~C.}\
  \bibnamefont {Leon}}, \bibinfo {author} {\bibfnamefont {C.~H.}\ \bibnamefont
  {Yang}}, \bibinfo {author} {\bibfnamefont {J.~C.~C.}\ \bibnamefont {Hwang}},
  \bibinfo {author} {\bibfnamefont {J.}~\bibnamefont {Camirand~Lemyre}},
  \bibinfo {author} {\bibfnamefont {T.}~\bibnamefont {Tanttu}}, \bibinfo
  {author} {\bibfnamefont {W.}~\bibnamefont {Huang}}, \bibinfo {author}
  {\bibfnamefont {J.~Y.}\ \bibnamefont {Huang}}, \bibinfo {author}
  {\bibfnamefont {F.~E.}\ \bibnamefont {Hudson}}, \bibinfo {author}
  {\bibfnamefont {K.~M.}\ \bibnamefont {Itoh}}, \bibinfo {author}
  {\bibfnamefont {A.}~\bibnamefont {Laucht}}, \bibinfo {author} {\bibfnamefont
  {M.}~\bibnamefont {Pioro-Ladri{\`e}re}}, \bibinfo {author} {\bibfnamefont
  {A.}~\bibnamefont {Saraiva}}, \ and\ \bibinfo {author} {\bibfnamefont
  {A.~S.}\ \bibnamefont {Dzurak}},\ }\href {\doibase
  10.1038/s41467-021-23437-w} {\bibfield  {journal} {\bibinfo  {journal}
  {Nature Communications}\ }\textbf {\bibinfo {volume} {12}},\ \bibinfo {pages}
  {3228} (\bibinfo {year} {2021})}\BibitemShut {NoStop}%
\bibitem [{\citenamefont {Corrigan}\ \emph {et~al.}(2021)\citenamefont
  {Corrigan}, \citenamefont {Dodson}, \citenamefont {Ercan}, \citenamefont
  {Abadillo-Uriel}, \citenamefont {Thorgrimsson}, \citenamefont {Knapp},
  \citenamefont {Holman}, \citenamefont {McJunkin}, \citenamefont {Neyens},
  \citenamefont {MacQuarrie}, \citenamefont {Foote}, \citenamefont {Edge},
  \citenamefont {Friesen}, \citenamefont {Coppersmith},\ and\ \citenamefont
  {Eriksson}}]{Corrigan:2021p127701}%
  \BibitemOpen
  \bibfield  {author} {\bibinfo {author} {\bibfnamefont {J.}~\bibnamefont
  {Corrigan}}, \bibinfo {author} {\bibfnamefont {J.~P.}\ \bibnamefont
  {Dodson}}, \bibinfo {author} {\bibfnamefont {H.~E.}\ \bibnamefont {Ercan}},
  \bibinfo {author} {\bibfnamefont {J.~C.}\ \bibnamefont {Abadillo-Uriel}},
  \bibinfo {author} {\bibfnamefont {B.}~\bibnamefont {Thorgrimsson}}, \bibinfo
  {author} {\bibfnamefont {T.~J.}\ \bibnamefont {Knapp}}, \bibinfo {author}
  {\bibfnamefont {N.}~\bibnamefont {Holman}}, \bibinfo {author} {\bibfnamefont
  {T.}~\bibnamefont {McJunkin}}, \bibinfo {author} {\bibfnamefont {S.~F.}\
  \bibnamefont {Neyens}}, \bibinfo {author} {\bibfnamefont {E.~R.}\
  \bibnamefont {MacQuarrie}}, \bibinfo {author} {\bibfnamefont {R.~H.}\
  \bibnamefont {Foote}}, \bibinfo {author} {\bibfnamefont {L.~F.}\ \bibnamefont
  {Edge}}, \bibinfo {author} {\bibfnamefont {M.}~\bibnamefont {Friesen}},
  \bibinfo {author} {\bibfnamefont {S.~N.}\ \bibnamefont {Coppersmith}}, \ and\
  \bibinfo {author} {\bibfnamefont {M.~A.}\ \bibnamefont {Eriksson}},\ }\href
  {\doibase 10.1103/PhysRevLett.127.127701} {\bibfield  {journal} {\bibinfo
  {journal} {Phys. Rev. Lett.}\ }\textbf {\bibinfo {volume} {127}},\ \bibinfo
  {pages} {127701} (\bibinfo {year} {2021})}\BibitemShut {NoStop}%
\bibitem [{\citenamefont {Ercan}\ \emph {et~al.}(2021)\citenamefont {Ercan},
  \citenamefont {Coppersmith},\ and\ \citenamefont
  {Friesen}}]{Ercan:2021p235302}%
  \BibitemOpen
  \bibfield  {author} {\bibinfo {author} {\bibfnamefont {H.~E.}\ \bibnamefont
  {Ercan}}, \bibinfo {author} {\bibfnamefont {S.~N.}\ \bibnamefont
  {Coppersmith}}, \ and\ \bibinfo {author} {\bibfnamefont {M.}~\bibnamefont
  {Friesen}},\ }\href {\doibase 10.1103/PhysRevB.104.235302} {\bibfield
  {journal} {\bibinfo  {journal} {Phys. Rev. B}\ }\textbf {\bibinfo {volume}
  {104}},\ \bibinfo {pages} {235302} (\bibinfo {year} {2021})}\BibitemShut
  {NoStop}%
\bibitem [{\citenamefont {Abadillo-Uriel}\ \emph {et~al.}(2021)\citenamefont
  {Abadillo-Uriel}, \citenamefont {Martinez}, \citenamefont {Filippone},\ and\
  \citenamefont {Niquet}}]{Abadillo-Uriel:2021p195305}%
  \BibitemOpen
  \bibfield  {author} {\bibinfo {author} {\bibfnamefont {J.~C.}\ \bibnamefont
  {Abadillo-Uriel}}, \bibinfo {author} {\bibfnamefont {B.}~\bibnamefont
  {Martinez}}, \bibinfo {author} {\bibfnamefont {M.}~\bibnamefont {Filippone}},
  \ and\ \bibinfo {author} {\bibfnamefont {Y.-M.}\ \bibnamefont {Niquet}},\
  }\href {\doibase 10.1103/PhysRevB.104.195305} {\bibfield  {journal} {\bibinfo
   {journal} {Phys. Rev. B}\ }\textbf {\bibinfo {volume} {104}},\ \bibinfo
  {pages} {195305} (\bibinfo {year} {2021})}\BibitemShut {NoStop}%
\bibitem [{\citenamefont {Yannouleas}\ and\ \citenamefont
  {Landman}(2022{\natexlab{a}})}]{Yannouleas:2022p21LT01}%
  \BibitemOpen
  \bibfield  {author} {\bibinfo {author} {\bibfnamefont {C.}~\bibnamefont
  {Yannouleas}}\ and\ \bibinfo {author} {\bibfnamefont {U.}~\bibnamefont
  {Landman}},\ }\href {\doibase 10.1088/1361-648x/ac5c28} {\bibfield  {journal}
  {\bibinfo  {journal} {Journal of Physics: Condensed Matter}\ }\textbf
  {\bibinfo {volume} {34}},\ \bibinfo {pages} {21LT01} (\bibinfo {year}
  {2022}{\natexlab{a}})}\BibitemShut {NoStop}%
\bibitem [{\citenamefont {Yannouleas}\ and\ \citenamefont
  {Landman}(2022{\natexlab{b}})}]{Yannouleas:2022p205302}%
  \BibitemOpen
  \bibfield  {author} {\bibinfo {author} {\bibfnamefont {C.}~\bibnamefont
  {Yannouleas}}\ and\ \bibinfo {author} {\bibfnamefont {U.}~\bibnamefont
  {Landman}},\ }\href {\doibase 10.1103/PhysRevB.105.205302} {\bibfield
  {journal} {\bibinfo  {journal} {Phys. Rev. B}\ }\textbf {\bibinfo {volume}
  {105}},\ \bibinfo {pages} {205302} (\bibinfo {year}
  {2022}{\natexlab{b}})}\BibitemShut {NoStop}%
\bibitem [{\citenamefont {Yannouleas}\ and\ \citenamefont
  {Landman}(2022{\natexlab{c}})}]{Yannouleas:2022preprint}%
  \BibitemOpen
  \bibfield  {author} {\bibinfo {author} {\bibfnamefont {C.}~\bibnamefont
  {Yannouleas}}\ and\ \bibinfo {author} {\bibfnamefont {U.}~\bibnamefont
  {Landman}},\ }\href@noop {} {\  (\bibinfo {year} {2022}{\natexlab{c}})},\
  \Eprint {http://arxiv.org/abs/2208.05626} {arXiv:2208.05626
  [cond-mat.mes-hall]} \BibitemShut {NoStop}%
\bibitem [{\citenamefont {Zwanenburg}\ \emph {et~al.}(2013)\citenamefont
  {Zwanenburg}, \citenamefont {Dzurak}, \citenamefont {Morello}, \citenamefont
  {Simmons}, \citenamefont {Hollenberg}, \citenamefont {Klimeck}, \citenamefont
  {Rogge}, \citenamefont {Coppersmith},\ and\ \citenamefont
  {Eriksson}}]{Zwanenburg:2013p961}%
  \BibitemOpen
  \bibfield  {author} {\bibinfo {author} {\bibfnamefont {F.~A.}\ \bibnamefont
  {Zwanenburg}}, \bibinfo {author} {\bibfnamefont {A.~S.}\ \bibnamefont
  {Dzurak}}, \bibinfo {author} {\bibfnamefont {A.}~\bibnamefont {Morello}},
  \bibinfo {author} {\bibfnamefont {M.~Y.}\ \bibnamefont {Simmons}}, \bibinfo
  {author} {\bibfnamefont {L.~C.~L.}\ \bibnamefont {Hollenberg}}, \bibinfo
  {author} {\bibfnamefont {G.}~\bibnamefont {Klimeck}}, \bibinfo {author}
  {\bibfnamefont {S.}~\bibnamefont {Rogge}}, \bibinfo {author} {\bibfnamefont
  {S.~N.}\ \bibnamefont {Coppersmith}}, \ and\ \bibinfo {author} {\bibfnamefont
  {M.~A.}\ \bibnamefont {Eriksson}},\ }\href {\doibase
  10.1103/RevModPhys.85.961} {\bibfield  {journal} {\bibinfo  {journal} {Rev.
  Mod. Phys.}\ }\textbf {\bibinfo {volume} {85}},\ \bibinfo {pages} {961}
  (\bibinfo {year} {2013})}\BibitemShut {NoStop}%
\bibitem [{\citenamefont {Croot}\ \emph {et~al.}(2020)\citenamefont {Croot},
  \citenamefont {Mi}, \citenamefont {Putz}, \citenamefont {Benito},
  \citenamefont {Borjans}, \citenamefont {Burkard},\ and\ \citenamefont
  {Petta}}]{Croot:2020p012006}%
  \BibitemOpen
  \bibfield  {author} {\bibinfo {author} {\bibfnamefont {X.}~\bibnamefont
  {Croot}}, \bibinfo {author} {\bibfnamefont {X.}~\bibnamefont {Mi}}, \bibinfo
  {author} {\bibfnamefont {S.}~\bibnamefont {Putz}}, \bibinfo {author}
  {\bibfnamefont {M.}~\bibnamefont {Benito}}, \bibinfo {author} {\bibfnamefont
  {F.}~\bibnamefont {Borjans}}, \bibinfo {author} {\bibfnamefont
  {G.}~\bibnamefont {Burkard}}, \ and\ \bibinfo {author} {\bibfnamefont
  {J.~R.}\ \bibnamefont {Petta}},\ }\href {\doibase
  10.1103/PhysRevResearch.2.012006} {\bibfield  {journal} {\bibinfo  {journal}
  {Phys. Rev. Research}\ }\textbf {\bibinfo {volume} {2}},\ \bibinfo {pages}
  {012006} (\bibinfo {year} {2020})}\BibitemShut {NoStop}%
\bibitem [{\citenamefont {Anderson}\ \emph {et~al.}(2022)\citenamefont
  {Anderson}, \citenamefont {Gyure}, \citenamefont {Quinn}, \citenamefont
  {Pan}, \citenamefont {Ross},\ and\ \citenamefont
  {Kiselev}}]{Anderson:2022p065123}%
  \BibitemOpen
  \bibfield  {author} {\bibinfo {author} {\bibfnamefont {C.~R.}\ \bibnamefont
  {Anderson}}, \bibinfo {author} {\bibfnamefont {M.~F.}\ \bibnamefont {Gyure}},
  \bibinfo {author} {\bibfnamefont {S.}~\bibnamefont {Quinn}}, \bibinfo
  {author} {\bibfnamefont {A.}~\bibnamefont {Pan}}, \bibinfo {author}
  {\bibfnamefont {R.~S.}\ \bibnamefont {Ross}}, \ and\ \bibinfo {author}
  {\bibfnamefont {A.~A.}\ \bibnamefont {Kiselev}},\ }\href {\doibase
  10.1063/5.0089350} {\bibfield  {journal} {\bibinfo  {journal} {AIP Advances}\
  }\textbf {\bibinfo {volume} {12}},\ \bibinfo {pages} {065123} (\bibinfo
  {year} {2022})}\BibitemShut {NoStop}%
\bibitem [{\citenamefont {Boykin}\ \emph
  {et~al.}(2004{\natexlab{a}})\citenamefont {Boykin}, \citenamefont {Klimeck},
  \citenamefont {Eriksson}, \citenamefont {Friesen}, \citenamefont
  {Coppersmith}, \citenamefont {von Allmen}, \citenamefont {Oyafuso},\ and\
  \citenamefont {Lee}}]{Boykin:2004p115}%
  \BibitemOpen
  \bibfield  {author} {\bibinfo {author} {\bibfnamefont {T.~B.}\ \bibnamefont
  {Boykin}}, \bibinfo {author} {\bibfnamefont {G.}~\bibnamefont {Klimeck}},
  \bibinfo {author} {\bibfnamefont {M.~A.}\ \bibnamefont {Eriksson}}, \bibinfo
  {author} {\bibfnamefont {M.}~\bibnamefont {Friesen}}, \bibinfo {author}
  {\bibfnamefont {S.~N.}\ \bibnamefont {Coppersmith}}, \bibinfo {author}
  {\bibfnamefont {P.}~\bibnamefont {von Allmen}}, \bibinfo {author}
  {\bibfnamefont {F.}~\bibnamefont {Oyafuso}}, \ and\ \bibinfo {author}
  {\bibfnamefont {S.}~\bibnamefont {Lee}},\ }\href {\doibase 10.1063/1.1637718}
  {\bibfield  {journal} {\bibinfo  {journal} {Appl. Phys. Lett.}\ }\textbf
  {\bibinfo {volume} {84}},\ \bibinfo {pages} {115} (\bibinfo {year}
  {2004}{\natexlab{a}})}\BibitemShut {NoStop}%
\bibitem [{\citenamefont {Boykin}\ \emph
  {et~al.}(2004{\natexlab{b}})\citenamefont {Boykin}, \citenamefont {Klimeck},
  \citenamefont {Friesen}, \citenamefont {Coppersmith}, \citenamefont
  {vonAllmen}, \citenamefont {Oyafuso},\ and\ \citenamefont
  {Lee}}]{Boykin:2004p165325}%
  \BibitemOpen
  \bibfield  {author} {\bibinfo {author} {\bibfnamefont {T.~B.}\ \bibnamefont
  {Boykin}}, \bibinfo {author} {\bibfnamefont {G.}~\bibnamefont {Klimeck}},
  \bibinfo {author} {\bibfnamefont {M.}~\bibnamefont {Friesen}}, \bibinfo
  {author} {\bibfnamefont {S.~N.}\ \bibnamefont {Coppersmith}}, \bibinfo
  {author} {\bibfnamefont {P.}~\bibnamefont {vonAllmen}}, \bibinfo {author}
  {\bibfnamefont {F.}~\bibnamefont {Oyafuso}}, \ and\ \bibinfo {author}
  {\bibfnamefont {S.}~\bibnamefont {Lee}},\ }\href {\doibase
  10.1103/PhysRevB.70.165325} {\bibfield  {journal} {\bibinfo  {journal} {Phys.
  Rev. B}\ }\textbf {\bibinfo {volume} {70}},\ \bibinfo {pages} {165325}
  (\bibinfo {year} {2004}{\natexlab{b}})}\BibitemShut {NoStop}%
\bibitem [{\citenamefont {Friesen}\ and\ \citenamefont
  {Coppersmith}(2010)}]{Friesen:2010p115324}%
  \BibitemOpen
  \bibfield  {author} {\bibinfo {author} {\bibfnamefont {M.}~\bibnamefont
  {Friesen}}\ and\ \bibinfo {author} {\bibfnamefont {S.~N.}\ \bibnamefont
  {Coppersmith}},\ }\href {\doibase 10.1103/PhysRevB.81.115324} {\bibfield
  {journal} {\bibinfo  {journal} {Phys. Rev. B}\ }\textbf {\bibinfo {volume}
  {81}},\ \bibinfo {pages} {115324} (\bibinfo {year} {2010})}\BibitemShut
  {NoStop}%
\bibitem [{\citenamefont {Abadillo-Uriel}\ \emph {et~al.}(2018)\citenamefont
  {Abadillo-Uriel}, \citenamefont {Thorgrimsson}, \citenamefont {Kim},
  \citenamefont {Smith}, \citenamefont {Simmons}, \citenamefont {Ward},
  \citenamefont {Foote}, \citenamefont {Corrigan}, \citenamefont {Savage},
  \citenamefont {Lagally}, \citenamefont {Calder\'on}, \citenamefont
  {Coppersmith}, \citenamefont {Eriksson},\ and\ \citenamefont
  {Friesen}}]{Abadillo-Uriel:2018p165438}%
  \BibitemOpen
  \bibfield  {author} {\bibinfo {author} {\bibfnamefont {J.~C.}\ \bibnamefont
  {Abadillo-Uriel}}, \bibinfo {author} {\bibfnamefont {B.}~\bibnamefont
  {Thorgrimsson}}, \bibinfo {author} {\bibfnamefont {D.}~\bibnamefont {Kim}},
  \bibinfo {author} {\bibfnamefont {L.~W.}\ \bibnamefont {Smith}}, \bibinfo
  {author} {\bibfnamefont {C.~B.}\ \bibnamefont {Simmons}}, \bibinfo {author}
  {\bibfnamefont {D.~R.}\ \bibnamefont {Ward}}, \bibinfo {author}
  {\bibfnamefont {R.~H.}\ \bibnamefont {Foote}}, \bibinfo {author}
  {\bibfnamefont {J.}~\bibnamefont {Corrigan}}, \bibinfo {author}
  {\bibfnamefont {D.~E.}\ \bibnamefont {Savage}}, \bibinfo {author}
  {\bibfnamefont {M.~G.}\ \bibnamefont {Lagally}}, \bibinfo {author}
  {\bibfnamefont {M.~J.}\ \bibnamefont {Calder\'on}}, \bibinfo {author}
  {\bibfnamefont {S.~N.}\ \bibnamefont {Coppersmith}}, \bibinfo {author}
  {\bibfnamefont {M.~A.}\ \bibnamefont {Eriksson}}, \ and\ \bibinfo {author}
  {\bibfnamefont {M.}~\bibnamefont {Friesen}},\ }\href {\doibase
  10.1103/PhysRevB.98.165438} {\bibfield  {journal} {\bibinfo  {journal} {Phys.
  Rev. B}\ }\textbf {\bibinfo {volume} {98}},\ \bibinfo {pages} {165438}
  (\bibinfo {year} {2018})}\BibitemShut {NoStop}%
\bibitem [{\citenamefont {Tokura}\ \emph {et~al.}(2006)\citenamefont {Tokura},
  \citenamefont {Van~der Wiel}, \citenamefont {Obata},\ and\ \citenamefont
  {Tarucha}}]{Tokura:2006p047202}%
  \BibitemOpen
  \bibfield  {author} {\bibinfo {author} {\bibfnamefont {Y.}~\bibnamefont
  {Tokura}}, \bibinfo {author} {\bibfnamefont {W.~G.}\ \bibnamefont {Van~der
  Wiel}}, \bibinfo {author} {\bibfnamefont {T.}~\bibnamefont {Obata}}, \ and\
  \bibinfo {author} {\bibfnamefont {S.}~\bibnamefont {Tarucha}},\ }\href
  {\doibase 10.1103/PhysRevLett.96.047202} {\bibfield  {journal} {\bibinfo
  {journal} {Phys. Rev. Lett.}\ }\textbf {\bibinfo {volume} {96}},\ \bibinfo
  {pages} {047202} (\bibinfo {year} {2006})}\BibitemShut {NoStop}%
\bibitem [{Note1()}]{Note1}%
  \BibitemOpen
  \bibinfo {note} {Appendix \ref {sec:appEDSR3e} presents a simplified
  calculation, where the perturbation theory is only in the first order in
  $b$.}\BibitemShut {Stop}%
\bibitem [{\citenamefont {Ruan}\ \emph {et~al.}(1995)\citenamefont {Ruan},
  \citenamefont {Liu}, \citenamefont {Bao},\ and\ \citenamefont
  {Zhang}}]{Ruan:1995p7942}%
  \BibitemOpen
  \bibfield  {author} {\bibinfo {author} {\bibfnamefont {W.~Y.}\ \bibnamefont
  {Ruan}}, \bibinfo {author} {\bibfnamefont {Y.~Y.}\ \bibnamefont {Liu}},
  \bibinfo {author} {\bibfnamefont {C.~G.}\ \bibnamefont {Bao}}, \ and\
  \bibinfo {author} {\bibfnamefont {Z.~Q.}\ \bibnamefont {Zhang}},\ }\href
  {\doibase 10.1103/PhysRevB.51.7942} {\bibfield  {journal} {\bibinfo
  {journal} {Phys. Rev. B}\ }\textbf {\bibinfo {volume} {51}},\ \bibinfo
  {pages} {7942} (\bibinfo {year} {1995})}\BibitemShut {NoStop}%
\bibitem [{\citenamefont {Huang}\ \emph {et~al.}(2017)\citenamefont {Huang},
  \citenamefont {Veldhorst}, \citenamefont {Zimmerman}, \citenamefont
  {Dzurak},\ and\ \citenamefont {Culcer}}]{Huang:2017p75403}%
  \BibitemOpen
  \bibfield  {author} {\bibinfo {author} {\bibfnamefont {W.}~\bibnamefont
  {Huang}}, \bibinfo {author} {\bibfnamefont {M.}~\bibnamefont {Veldhorst}},
  \bibinfo {author} {\bibfnamefont {N.~M.}\ \bibnamefont {Zimmerman}}, \bibinfo
  {author} {\bibfnamefont {A.~S.}\ \bibnamefont {Dzurak}}, \ and\ \bibinfo
  {author} {\bibfnamefont {D.}~\bibnamefont {Culcer}},\ }\href {\doibase
  10.1103/PhysRevB.95.075403} {\bibfield  {journal} {\bibinfo  {journal} {Phys.
  Rev. B}\ }\textbf {\bibinfo {volume} {95}},\ \bibinfo {pages} {075403}
  (\bibinfo {year} {2017})}\BibitemShut {NoStop}%
\bibitem [{Note2()}]{Note2}%
  \BibitemOpen
  \bibinfo {note} {Valley physics can be turned off by simply replacing
  $H_K^{TB}$ with -($\hbar ^2/2m_l)\partial ^2/\partial z^2$}\BibitemShut
  {NoStop}%
\bibitem [{\citenamefont {Yang}\ \emph {et~al.}(2017)\citenamefont {Yang},
  \citenamefont {Coppersmith},\ and\ \citenamefont
  {Friesen}}]{Yang:2017p062321}%
  \BibitemOpen
  \bibfield  {author} {\bibinfo {author} {\bibfnamefont {Y.-C.}\ \bibnamefont
  {Yang}}, \bibinfo {author} {\bibfnamefont {S.~N.}\ \bibnamefont
  {Coppersmith}}, \ and\ \bibinfo {author} {\bibfnamefont {M.}~\bibnamefont
  {Friesen}},\ }\href {\doibase 10.1103/PhysRevA.95.062321} {\bibfield
  {journal} {\bibinfo  {journal} {Physical Review A}\ }\textbf {\bibinfo
  {volume} {95}},\ \bibinfo {pages} {062321} (\bibinfo {year}
  {2017})}\BibitemShut {NoStop}%
\bibitem [{\citenamefont {Rashba}(2008)}]{Rashba:2008p195302}%
  \BibitemOpen
  \bibfield  {author} {\bibinfo {author} {\bibfnamefont {E.~I.}\ \bibnamefont
  {Rashba}},\ }\href {\doibase 10.1103/PhysRevB.78.195302} {\bibfield
  {journal} {\bibinfo  {journal} {Phys. Rev. B}\ }\textbf {\bibinfo {volume}
  {78}},\ \bibinfo {pages} {195302} (\bibinfo {year} {2008})}\BibitemShut
  {NoStop}%
\end{thebibliography}%

\end{document}